\documentclass[12pt,reqno]{article}
\usepackage{amsmath, amsthm, amssymb,epsfig,alltt}

\newcommand{\EQA}[1]{\begin{eqnarray}#1\end{eqnarray} }

\newcommand{\gbar}{\bar{g} }
\newcommand{\ditau}{\partial_\tau }
\newcommand{\BDket}{\left|B,D\right\rangle }

\newcommand{\twobytwo}[2]{\left[\begin{matrix} #1 \cr #2 \cr \end{matrix}\right] }

\begin{document}


\title{Critical Boundary Sine-Gordon Revisited }

\author{M.~Hasselfield$^1$, Taejin Lee$^{1,2,3}$,
G.W.~Semenoff$^{1 ,4 }$ and P.C.E.~Stamp$^1$\\~~\\ $^1$ Pacific
Institute for Theoretical Physics\\ and \\Department of Physics and
Astronomy, University of British Columbia\\ 6224 Agricultural Road,
Vancouver, British Columbia V6T 1Z1, Canada \\~~\\$^2$ Department of
Physics, Kangwon University, Chuncheon 200-701 Korea\\~~\\$^3$ Asia
Pacific Center for Theoretical Physics, Pohang 790-784 Korea
\\~~\\$^4$Institute des Hautes \'Etudes Scientifiques\\Le
Bois-Marie 35, F-91440 Bures-sur-Yvette, France
}

\maketitle

\centerline{\bf Abstract}

We revisit the exact solution of the two space-time dimensional
quantum field theory of a free massless boson with a periodic
boundary interaction and self-dual period. We analyze the model by
using a mapping to free fermions with a boundary mass term
originally suggested in ref.~\cite{Polchinski:1994my}.  We find that
the entire SL(2,C) family of boundary states of a single boson are
boundary sine-Gordon states and we derive a simple explicit
expression for the boundary state in fermion variables and as a
function of sine-Gordon coupling constants.  We use this expression
to compute the partition function. We observe that the solution of
the model has a strong-weak coupling generalization of T-duality. We
then examine a class of recently discovered conformal boundary
states for compact bosons with radii which are rational numbers
times the self-dual radius. These have simple expression in fermion
variables.  We postulate sine-Gordon-like field theories with
discrete gauge symmetries for which they are the appropriate
boundary states.

\vskip 2cm

\section{Introduction and Summary}

Boundary conformal field theory consisting of a single boson field
is of interest in a wide array of contexts. In condensed matter
physics it describes the dissipative quantum mechanics of a particle
in a one-dimensional periodic potential
\cite{schmid,Guinea,fisher,Callan:1989mm} and elaborations of it in
Josephson junction arrays \cite{larkin,fazio,sodano} and the
dissipative Hofstadter problem \cite{Callan:1991da,Callan:1992vy}.
It also arises in analysis of the Kondo problem
\cite{Affleck:1990iv,Affleck:1990by}, the study of one-dimensional
conductors \cite{Kane:1992}, tunneling between Hall edge states
\cite{kane2}, and junctions of quantum wires~\cite{Oshikawa:2005fh}.
In string theory, boundary conformal field theories are solutions of
classical open string field theory and those with boundary operators
describe open strings in background
fields~\cite{Witten:1992qy,Shatashvili:1993kk,Shatashvili:1993ps}.
In particular, a marginal, periodic boundary interaction which is
termed the ``rolling tachyon'' gives a description of the process of
tachyon condensation in string theories with unstable D-branes
\cite{Sen:2002nu}. The relationships between all of these theories
are not always trivial, and in this Paper we revisit the boundary
Sine-Gordon model from a string theory perspective. We shall see
that our results clearly have interesting implications for some of
the related models (particularly the condensed matter ones), but to
properly elaborate on these is beyond the scope of the present work.

Recently, two of the authors have discussed the rolling tachyon
boundary state by fermionizing the rolling tachyon boundary
conformal field theory and then finding the boundary states using
fermion variables~\cite{Lee:2005ge}. Here we shall extend that work,
which constructed the boundary state for the half-brane, to study
the full set of conformal boundary states which can be constructed
for a theory with a single free boson in the bulk. These states are
parameterized by SL(2,C) group elements and we find a simple mapping
of the SL(2,C) matrices onto the coupling constants of the boundary
sine-Gordon theory.  The fermion representation of this problem has
some advantages which we exploit, particularly for constructing
boundary states and computing partition functions. In addition, the
SL(2,C) symmetry and Kac-Moody algebra which appear at the self-dual
radius has a simple linear representation in terms of fermion
variables.  In this language, different boundary conditions are
simply related by global SL(2,C) transformations of the boundary
states. In particular, T-duality, which interchanges Neumann and
Dirichlet coordinates, is a simple chiral rotation. We shall use
this observation to show that a generalization of T-duality persists
when the critical boundary sine-Gordon interaction is turned on and
appears as a strong coupling-weak coupling duality of the critical
theory.

We shall discuss conformal boundary states which have recently been
shown to appear for compact bosons with rational radii
\cite{Gaberdiel:2001zq}. These also have a simple description in
fermion variables as projections of the SL(2,C) boundary states
onto a sectors of the theory with some fixed fermion charges.
These projections can be viewed as Gauss' law constraints for a
discrete gauge symmetry. We use our mapping of boundary states
onto boundary sine-Gordon models to suggest a gauged boundary
sine-Gordon field theory for which these are the appropriate
boundary states.

Our central result is an improved understanding of how the three
parameters $(g,\bar g, A)$ in the boundary sine-Gordon action of
eq.~(\ref{bosonaction}) below correspond to the three marginal
boundary deformations of $c=1$ conformal field theories
\cite{Recknagel:1998ih} and how they are encoded in boundary states
when the boundary states are written in fermion variables. We also
find that the fermion representation gives a solution of the theory
which has a different dependence on the sine-Gordon coupling
constants than the one which was found in the bosonic approach and
which is used extensively in the literature on tachyon condensation.
(See for example the scalar amplitude in eq.(\ref{1}) below compared
with appropriate Wick rotation with eq.~(4.10) in
ref.~\cite{Sen:2002nu}.) The dependence of the solution of a field
theory on its bare couplings is generally not universal, it is a
matter of the detailed definition of the theory, including cutoff
and renormalization procedures. However, knowledge of how the
solution of a theory depends on the bare coupling constants can be
important in condensed matter applications where the physical input
to the construction of a micro-electronic device, for example, are
the bare parameters. Our solution presents one such definition which
is an alternative to the other known ones, and which we believe is
natural.

\subsection{Critical Boundary Sine-Gordon Theory}

Consider the theory of a single boson field on a strip with time
coordinate $t\in (-\infty,\infty)$, space $s\in (0,\pi)$ and with
action
\begin{equation}\label{bosonaction}
S=\frac{1}{4\pi}\int_{-\infty}^\infty dt \int_0^\pi ds~\left(
\partial_tX^2-\partial_s X^2 \right) -\int_{-\infty}^\infty dt \left(
\frac{g}{2}e^{iX(t,0)}+\frac{\bar
g}{2}e^{-iX(t,0)}+A\partial_tX(t,0)\right)
\end{equation}
This is a free field theory in the bulk of the strip, with equation
of motion  $$\left(
\partial_t^2-\partial_s^2\right)X(t,s)=0$$ and is an interacting field theory by virtue of
the nonlinear boundary condition at $s=0$,
\begin{equation}
-\partial_s X(t,0)+i\frac{g}{2}e^{iX(t,0)}-i\frac{\bar
g}{2}e^{-iX(t,0)} = 0 \label{bc}\end{equation}  At the other
boundary, $s=\pi$, we will consider an array of possible boundary
conditions: the Dirichlet condition, $\partial_t X(t,\pi)=0$, the
Neumann condition $\partial_sX(t,\pi)=0$ and the addition of a
periodic boundary operator leading to a condition of type
(\ref{bc}), perhaps with different parameters.

We will allow $g$ and $\bar g$ to be independent complex numbers,
not necessarily complex conjugates of each other. Of course, for
the Hamiltonian to be Hermitian would require the special case
$\bar g=g^*$. The boundary interaction potential is known to be an
exactly marginal operator and (\ref{bosonaction}) is a conformal
field theory for all values of $g$ and $\bar g$.

This boundary conformal field theory is tuned to lie at the critical
point which separates two distinct behaviors of a one-dimensional
boson with a periodic boundary potential. In one regime, the boundary
potential would be a relevant operator. Its effect would be to
localize the boundary value of $X$ near the minima of the potential.
There, the coupling constant in the boundary interaction would have a
non-zero beta function.  The renormalization group flow would be
toward the conformal invariant Dirichlet boundary condition which
fixes the boundary $X$ at a particular location.

In the other regime, the boundary potential would be an irrelevant
operator. The boundary value of $X$ would be mobile and would
fluctuate amongst the minima of the periodic potential.  In that case,
the renormalization group flow of the coupling constant would be
toward the translation invariant and conformal invariant Neumann
boundary condition where complete translation invariance would be
restored.

In the boundary conformal field theory (\ref{bosonaction}), which
sits between these two phases, the boundary potential is an exactly
marginal operator.  This exactly solvable theory describes a family
of critical states which, as we vary the constants $g,\bar g$,
interpolate between the two kinds of behavior. $X$ becomes
delocalized when $g\sim 0\sim\bar g$, where the boundary condition
reverts to Neumann.  It is completely localized when the coupling
constant is increased to $\pi|g|=1=\pi|\bar g|$ where the boundary
state becomes Dirichlet and indeed pins the boundary value of $X$ at
the minima of the potential.  When $\pi|g|>1, \pi|\bar g|>1$ the
theory is apparently non-unitary.

The last, topological term in the action (\ref{bosonaction}) does
not influence the equations of motion or the boundary condition.
\footnote{The other possible simple  boundary term which one might
think of adding is $\int dt
\partial_sX(t,0)$.  This term could be eliminated by using the boundary
condition (\ref{bc}), and re-defining the coupling constants $g$ and
$\bar g$.} It does not destroy the conformal symmetry of the model
but we shall see that the spectrum generally depends on it. (It was
shown in refs.~\cite{Ahn:2003ns,Ahn:2005fj} that the conformal
dimensions of operators can depend on such a term.) For a
non-compact boson, it can be interpreted as a Bloch wave-number. To
see this, first consider the case with $A=0$ and the boson $X(t,s)$
taking values on the entire real line. The potential is periodic. We
assume that the boundary condition at $s=\pi$ is also periodic. Then
the translation $X(t,s)\to X(t,s)+2\pi$ is a discrete symmetry. The
eigenstates of the Hamiltonian must carry a representation of the
symmetry group. Since the group is Abelian, irreducible
representations are one-dimensional. Therefore a wave-functional
eigenstate of the Hamiltonian must transform as
\begin{equation}\label{phase1}
\Psi_E[X+2\pi ]= e^{-2\pi i A }\Psi_E[X]
\end{equation}
where $A$ takes values between zero and one. We can fix $A$, by
considering all states which transform by the same phase, $e^{-2\pi
iA}$. This is the content of Bloch's theorem and the phase is called
the Bloch wave-number.  In this sector of the theory, we can pass to
strictly periodic wave-functionals by a canonical transformation,
which acts on the wave-functional by a unitary operator.  The
operator is formed by exponentiating the generator of the
transformation,
\begin{equation}\label{bloch}
\Psi_E[X] = e^{-i AX(0,0)}\tilde\Psi_E[X] ~~,~~
\tilde\Psi_E[X+2\pi]=\tilde\Psi_E[X]
\end{equation}
Under such a canonical transformation, the Lagrangian is replaced
by itself plus the total time derivative of the generating
function,
\begin{equation}
S=\int dt L \to \int dt\left\{
L-\tfrac{d}{dt}\left[AX(t,0)\right]\right\}
\end{equation}
This restores the topological parameter $A$ in the action. In
solid-sate physics the functions $\Psi_E[X]$ are known as Wannier
functions. We see that the parameter $A$ coincides with the Bloch
wave-number of the boson field on its target space. Once it appears
explicitly in the action, when we compute the partition function or
the correlators of periodic operators, the boson field can be
treated as if it were compact, that is, as if it were subject to the
periodic identification $X(t,s)\sim X(t,s)+2\pi$.

If the boson were indeed compact, no physical observable,
including matrix elements between different wave-functions can be
changed by replacing $X$ by $X+2\pi$.   In this case, all
wave-functions must have the behavior (\ref{phase1}) with the same
parameter $A$.  In this case, we would simply leave $A$ as a
freely adjustable parameter of the theory.

For a non-compact boson, we would compute the partition function
or the correlator of periodic operators with $A$ fixed as if we
were computing the same object for a compact boson. Then we would
integrate the partition function or correlator over the allowed
values of $A$, from zero to one.

The periodicity of the potential is also compatible with a boson
which is identified with a period larger than the basic one, say
$X\sim X+2\pi N$.  To obtain this case, rather than integrating over
$A$, we would write $A=n/N+a$ and sum over $n=0,1,...,N-1$. The
theory would then be parameterized by another angle with
$a\in[~0,1/N)$. In this way one would develop a 'multiple band'
theory for this periodic potential- this in principle is quite
complicated because of transitions between the bands.

In string theory $A$ is interpreted as a constant gauge potential
on the target space, this is the origin of the terminology
``Wilson line''.

\subsection{Boundary versus bulk operators}

We will apply the boundary state technique to computation of the
partition function.  We will also find the boundary state in terms
of fermion variables.    A rough idea of the problem at hand can
be gotten by considering the scaling dimensions of the boundary
operator. Consider a free scalar field $X$ with Neumann boundary
conditions. When $:e^{iX}:$ is an operator living in the bulk of a
two-dimensional space it has conformal dimensions $(1/4,1/4)$,
whereas, when it is an operator on the boundary it has dimensions
$(1/2,1/2)$. The difference arises from the Neumann boundary
condition which fuses the boundary operator to its image, thereby
doubling its dimension.

We shall want to construct boundary states which are coherent
states of the bulk operators which satisfy the boundary condition.
We shall therefore need the boundary potential to be a fermion
bilinear in the bulk as well as on the boundary. A fermion
bilinear operator like $\psi_L^{\dagger}\psi_R$ has dimension
$(1/2,1/2)$ both in the bulk and on the boundary.  The boundary
dimension $(1/2,1/2)$ of $:e^{iX}:$ is just right to be a fermion
bilinear on the boundary, but its $(1/4,1/4)$ in the bulk is wrong
for a fermion bilinear, it cannot be one in the bulk.

The trick for getting around this difficulty was originally
applied to this model by Polchinski and Thorlacius
\cite{Polchinski:1994my} who introduced a second free boson $Y$
with Dirichlet boundary conditions.  Then $:e^{iY}:$ is a bulk
operator of dimension $(1/4,1/4)$ and a boundary operator of
dimension $(0,0)$. The composite $:e^{i(X+Y)}:$ has precisely the
property that we want, it has dimension $(1/2,1/2)$ both in the
bulk and on the boundary and can therefore be a fermion bilinear
in both environments. Because of the Dirichlet boundary condition,
$Y=0$ on the boundary, therefore, up to a multiplicative constant,
$:e^{i(X+Y)}:$ coincides with the operator of the boundary
potential, $:e^{iX}:$, at the boundary.

 Because $Y$ is a free boson with a simple boundary condition, its
 dynamics are exactly solvable and its contribution can easily be identified and discarded
 when computing the partition function or any correlators involving $X$ alone.
 Using this trick will allow us to fermionize the boundary interaction and to
construct boundary states. Most of the analysis of the present paper
centers on applying this idea to the boundary sine-Gordon theory.

\subsection{Summary of partition functions}

\subsubsection{Annulus amplitude}

 One of our main results is the general formula for
the partition function.  Consider the situation where there is a
boundary sine-Gordon interaction similar to the one in
eqn.(\ref{bosonaction}) on each of the two boundaries with
parameters $(g_i,\bar g_i, A_i)$ and $i=1,2$. In the path integral
representation of the partition function, the time is Euclidean and
is periodic, so the space-time has the topology of an annulus.

If $H$ is the Hamiltonian of the field theory in
(\ref{bosonaction}), we shall show that the partition function,
defined as $Z_{B_1B_2}[\beta]={\rm Tr}e^{-\beta H}$ is given by
the expression
\begin{equation}\label{theanswer} Z_{B_1B_2}[\beta] = \int_0^1
dA~e^{\beta/24}\sum_{n\in {\cal Z}} e^{-\beta(n+\delta)^2}
\prod_{k=1}^\infty \frac{1}{1-e^{-\beta k}}
\end{equation}
where
\begin{equation}\label{theanswer2}
\delta[A,g,\bar g]
=\frac{1}{2\pi}\cos^{-1}\left\{ \cos (2\pi A)\sqrt{
1-\pi^2g_1\bar g_1}\sqrt{1-\pi^2g_2\bar g_2}+\frac{\pi^2}{2}\left(
g_1\bar g_2+\bar g_1 g_2\right)\right\}
\end{equation}
Here $A=A_1-A_2$.

The result in (\ref{theanswer}) is presented for a non-compact
 boson.  If the boson were compact, with the self-dual radius, we
 would omit the integral over $A$, which would then remain as a
 parameter of the theory.  In that case, by letting the parameters
 $(g_i,\bar g_i,A_i)$ be complex numbers, and by varying them we sweep
 over the full set of SL(2,C) conformal boundary states, and
 therefore the full set of boundary conformal field theories with
 the periodicity $X\sim X+2\pi$.

 From (\ref{theanswer}) and (\ref{theanswer2}) we see that the ground
state energy is
 \begin{equation}\label{groundstateenergy}
 E_0= \delta^2  - \frac{1}{24}
 \end{equation}
where we choose the branch of delta with smallest absolute value.
This energy is generally real when $\bar g_i = g_i^*$ and when $A$ is
real.  For a compact boson it depends on the Wilson line parameter $A$
in a simple way.  If the boson is non-compact, we should instead think
of (\ref{groundstateenergy}) as an energy band, which varies over the
energies in the band as we vary $A$ over its range. For an example,
see Fig.~1.

The energies of excited states are
\begin{equation}\label{energies}
E=\left( n+\delta\right)^2 + N - \frac{1}{24}
\end{equation}
where $n$ is any integer and $N$ is a non-negative integer.  The
degeneracy of each state is given by $p[N]$, the number of
partitions of $N$ (where we define $p[0]=1$).

 Other partition functions are easily obtained as limits of
 (\ref{theanswer}).  The boundary condition reduces to a Neumann
 condition if we put $g=0=\bar g$ and a Dirichlet condition if we put
 $g = \bar g^*$ and $|g|= 1/\pi$. In the second case, the partition
 function no longer depends on $A$ and the phase of $-g$ gives the
 D-brane position on the circle, or if the theory is non-compact, the
 positions of an infinite periodic array of D-branes.

\subsubsection{Disc amplitude}

For some applications, for example the dissipative particle, path
integral representation of the partition function has the field
theory living on a Euclidean space with the geometry a
disc~\cite{Callan:1989mm}. The boundary interaction lies on the
boundary of the disc.  The most accessible quantity in the boundary
state formalism is the expectation value of a vertex operator such
as $e^{ik_LX_L+ik_RX_R}$ which is inserted into the bulk of the
disc. Expectation values of these operators  are computed explicitly
in Section 5 for all values of $(k_L,k_R)$ and have a very simple
expression.  With them, we can find the inner product of the bulk
boson position eigenstate  $|X_{0L},X_{0R}>$ and the boundary state
$$ |B> = |X_{0L},X_{0R}><X_{0L},X_{0R}|B>+ ~(~{\rm oscillators}~)~ |X_{0L},X_{0R}> $$ For a
non-compact boson, the result is
\begin{equation}
2^{\tfrac{1}{4}}<X_0=X_{0L}+X_{0R}|B> = \frac{1}{2\pi}\left[
\frac{1}{1+\pi g e^{iX_0}}+ \frac{1}{1+\pi \bar g e^{-iX_0}} -
1\right] \label{1}
\end{equation}
For a compact boson, with the identification $X\sim X+2\pi$, the result
is similar, but also retains a contribution of the wrapped states,
\begin{eqnarray}
2^{\tfrac{1}{4}}<X_{0L},X_{0R}|B> &=& \frac{1}{2\pi}\left[
\frac{1}{1+\pi g e^{iX_0}}+ \frac{1}{1+\pi \bar g e^{-iX_0}} +\frac{
1}{1-\sqrt{1-\pi^2 g\bar g}e^{i\hat X_0+2\pi i A}}\right.
\nonumber\\
&& \qquad \left.+ \frac{
1}{1-\sqrt{1-\pi^2 g\bar g}e^{-i\hat X_0-2\pi i A}}-3\right] \label{2}
\end{eqnarray}
where $X_0=X_{0L}+X_{0R}$ is position on the circle and $\hat
X_0=X_{0L}-X_{0R}$ is position on the dual circle. Note that $\hat
X_0$ is redundant with the Wilson line $A$. In Section 4, we will
discuss why this is expected.

For boundary states where the boson is compact with
a rational radius, $X\sim X+2\pi M/N$,
\begin{eqnarray}
2^{\tfrac{1}{4}}<X_{0L},X_{0R}|B> = \frac{1}{2\pi}\left[
\frac{1}{1+(\pi g)^M e^{iMX_0}}+ \frac{1}{1+(\pi \bar g)^M
e^{-iMX_0}} +~~~~~~~\right. \nonumber
\\ \left.  +\frac{ 1}{1-(1-\pi^2 g\bar g)^{N/2}e^{iN\hat X_02\pi
iN A}}+ \frac{ 1}{1-(1-\pi^2 g\bar g)^{N/2}e^{-i N\hat X_0-2\pi i
NA}}-3\right] \label{3}\end{eqnarray} Here, the exponents $M$ and
$N$ arise from the fact that these boundary states are very similar
to the one which gives (\ref{2}), the main difference being that
states with unwanted momentum and wrapping numbers have been
eliminated.  This cull of states leads to a similar cull of powers
of $e^{iX_0}$ and $e^{i\hat X_0}$ leaving only multiples of $M$ and
$N$ respectively.

\subsubsection{Duality}

Eqs.~(\ref{2}) and (\ref{3}) have an interesting apparent
symmetry.  If we make the interchange of coordinate and dual
coordinate,
$$
X_0 ~~ \leftrightarrow ~~ \hat X_0
$$
that is normally made to implement T-duality, and we also make the
replacements
\begin{subequations}
\label{generallabel}
\begin{eqnarray}
\label{td1}\pi g  &\leftrightarrow &
 -\sqrt{1-\pi^2 g \bar g}e^{2\pi i A} \\
\label{td2}\pi \bar g &\leftrightarrow & -\sqrt{1-\pi^2 g \bar
g}e^{-2\pi i A}
\end{eqnarray}
\end{subequations}
we see that the amplitudes in
eqs.~(\ref{2}) and ~(\ref{3}) retain the same form. In fact, the
entire energy spectrum (\ref{energies}) is invariant. This
replacement -- which at the operator level is $(X_L,X_R)\to
(X_L,-X_R)$ -- is a symmetry of the field theory which can easily
be seen in the boundary state formalism.  It is a generalization
of T-duality and it maps to each other the weak coupling and
strong coupling regimes of the conformal field theory.

\section{Boundary states}

We shall apply the boundary state technique to the problem of
computing the partition function,
$$
Z [\beta] = {\rm Tr} e^{-\beta H}
$$
where $H$ is the Hamiltonian of the field theory in
(\ref{bosonaction}). When this partition function is computed using
the path integral formulation, the action has periodically
identified Euclidean time $\tau\in(0,\beta)$ so that the space-time
geometry is a Euclidean cylinder.

The boundary state technique interchanges the space and time
coordinates on the cylinder to compute the same partition function
as a matrix element of the Euclidean time evolution operator for a
free boson on a cylinder between boundary states which encode all of
the information about the boundary conditions,
\begin{equation}\label{bspart}
Z[\beta]= <B_1|e^{-\frac{2\pi^2}{\beta} \hat H}|B_2>
\end{equation}
Now, the space variable is periodic, $\sigma\sim\sigma+\beta$ and
the time variable goes between the boundaries of the cylinder
$0\leq\tau\leq\pi$.  In the following we shall use conformal
invariance to re-scale the dimensions to match usual conventions for
open and closed strings where $0\leq\sigma<2\pi$. We have
anticipated this re-scaling in the Euclidean time parameter,
$\alpha=\frac{2\pi^2}{\beta}$, in (\ref{bspart}). The Hamiltonian
$\hat H$ in (\ref{bspart}) is that of the free boson on a Euclidean
cylinder, which has action
\begin{equation}
\hat S=\frac{1}{4\pi}\int_{-\infty}^\infty d\tau \int_0^{2\pi}
d\sigma\left(
\partial_\tau X^2+\partial_\sigma X^2\right)
\label{euclideanboseaction}\end{equation} with
$X(\tau,2\pi)=X(\tau,0)$. The conformal transformation
$z=e^{\tau+i\sigma}$, $\bar z = e^{\tau-i\sigma}$ maps the cylinder
to the complex plane with boundaries of the cylinder mapped to the
origin and the circle at infinity. Radial quantization of
(\ref{euclideanboseaction})  has the equation of motion
\begin{equation}\label{eqnmotion}\frac{\partial}{\partial\bar z}\frac{
\partial}{\partial z} X(z,\bar z)=0\end{equation}
which is solved by $X(z,\bar z)=X_L(z)+X_R(\bar z)$ with the mode
expansion
$$ X_L(z)=\frac{1}{\sqrt{2}}\left( x_L-ip_L\ln
z+i\sum_{n\neq 0}\frac{\alpha_n}{n}z^{-n}\right)~,~ X_R(\bar
z)=\frac{1}{\sqrt{2}}\left( x_R-ip_R\ln \bar z+i\sum_{n\neq
0}\frac{\tilde\alpha_n}{n}{\bar z}^{-n}\right)$$ where the
non-vanishing commutators $\left[ x_L,p_L\right]=i$, $\left[
x_R,p_R\right]=i$, $\left[ \alpha_m,\alpha_n\right]=m\delta_{m+n}$,
$\left[\tilde\alpha_m,\tilde \alpha_n\right]=m\delta_{m+n}$
represent the canonical commutator in Euclidean time,
$$\left[ X(0,\sigma),\partial_\tau
X(0,\sigma')\right]=2\pi\delta(\sigma-\sigma')$$   The Hamiltonian
is \begin{equation}\label{bosehamiltonian} \hat
H=\frac{1}{2}p_L^2+\frac{1}{2}p_R^2+\sum_{n=1}^\infty
\left(\alpha_{-n}\alpha_n+\tilde\alpha_{-n}\tilde\alpha_n\right)-\frac{1}{12}
\end{equation}
The vacuum $|p_L,p_R>$ is an eigenstate of $p_L$ and $p_R$ and is
annihilated by all positively moded oscillators $\alpha_n$ and
$\tilde\alpha_n$ with $n>0$. For a compact boson with the same
period as the potential, the momenta are quantized,
\begin{equation}X\sim X+2\pi~\to~ \frac{1}{\sqrt{2}}\left(p_L+p_R\right)={\rm integer}
~~,~~\frac{1}{\sqrt{2}}\left(p_L-p_R\right)={\rm
integer}\label{radius}\end{equation}

The boundary states are squeezed states which are annihilated by
the operator which would be the boundary condition for the field
theory on the strip (with $\sigma$ and $\tau$ interchanged).
Examples are the Neumann $|N>$and Dirichlet $|D>$ states which
obey the Neumann and Dirichlet boundary conditions \footnote{ In
the following, we shall sometimes use the Neuman and Dirichlet
boundary conditions in the stronger forms
\begin{equation}\label{strongform}\left(X_L(0,\sigma)-X_R(0,\sigma)\right)|N>=0
~~,~~\left( X_L(0,\sigma)+X_R(0,\sigma)\right)|D>=0\end{equation}
These conditions are solved by the boundary states $|N>$ and $|D>$
in (\ref{bstates1}) and (\ref{bstates2}) with $A=0$. }
\begin{eqnarray}\label{NandD}
\left( \partial_\tau X_L(0,\sigma)+\partial_\tau
X_R(0,\sigma)\right)|N>=0 ~~\left(\partial_\tau
X_L(0,\sigma)-\partial_\tau X_R(0,\sigma)\right)|D>=0
\end{eqnarray}
respectively.  These are solved by\footnote{The awkward factor of
$2^{-\tfrac{1}{4}}$ in the normalization of these states arises
from the need to produce the correct partition function for the
boson theory on the strip.}
\begin{subequations}
\label{generallabel}
\begin{eqnarray}|N> &=& 2^{-\tfrac{1}{4}}\prod_{n>0}e^{-\frac{1}{n}\alpha_{-n}\tilde
    \alpha_{-n}}\sum_{p_L}e^{-2\pi iA\sqrt{2}p_L}|p_L,-p_L>  \label{bstates1} \\
  |D> &=& 2^{-\tfrac{1}{4}}\prod_{n>0}e^{\frac{1}{n}\alpha_{-n}\tilde
    \alpha_{-n}}\sum_{p_L}e^{-2\pi
    iA\sqrt{2}p_L}|p_L,p_L>. \label{bstates2}
\end{eqnarray}
\end{subequations}
These states
are related to each other by T-duality, $(X_L,X_R)\to (X_L,-X_R)$,
which interchanges Neuman and Dirichelt boundary conditions. $A$
is a parameter of the solutions. It is the Wilson line for the
Neuman state, analogous to the parameter in (\ref{bosonaction}).
For the Dirichlet state, it is the position of the D-brane,
$X(0,\sigma)|D> = -2\pi (A~{\rm mod}~n)|D>$.

The boundary state which obeys
\begin{equation}\left(-\partial_\tau
X(0,\sigma)+i\frac{g}{2}e^{iX(0,\sigma)}-i\frac{\bar
g}{2}e^{-iX(0,\sigma)}\right)|B> = 0\label{bsgbs}\end{equation} is
also known explicitly. It has been constructed in
ref.~\cite{Callan:1994ub} using the level-1 SU(2) Kac-Moody
algebra which appears at the self-dual radius
(\ref{radius}).\footnote{The same current algebra is also the
vertex operator algebra of the discrete states of special `primary
operators of non-compact bosons \cite{Klebanov:1991hx} which, if
we do not compactify the boson, still play a special role. See
ref.~\cite{Callan:1994ub} for some discussion of this point.} More
recently, a large class of correlation functions has been computed
\cite{Kristjansson:2004ny}. Almost simultaneously with the
appearance of ref.~\cite{Callan:1994ub}, the model
(\ref{bosonaction}) (with $\bar g=g^*$) was also solved in
ref.~\cite{Polchinski:1994my} by mapping it onto a theory of free
fermions with a boundary mass operator. They did not construct
boundary states but did compute the partition function and
boundary S-matrix and their results agreed with those of
ref.~\cite{Callan:1994ub}. A limit of this model where $\bar g=0$
was solved by construction of the boundary state in fermion
variables in ref.~\cite{Lee:2005ge}. In the following, we will
employ the techniques of ref.~\cite{Lee:2005ge} to find the
boundary state $|B>$ in terms of fermion variables and compute the
partition function $Z[\beta]$ as a function of $g,\bar g$ and $A$
for various combinations of boundary conditions.

\subsection{ Fermionization at the self-dual radius}

Let us re-write the problem of finding the boundary state $|B>$ in
terms of fermion variables.  Following
refs.~\cite{Polchinski:1994my} and \cite{Lee:2005ge}, the first step
is to double the degrees of freedom by introducing another boson
field $Y$ which obeys a Dirichlet boundary condition on all
boundaries. The Euclidean action becomes
\begin{equation}\label{doublebosonaction}
S=\frac{1}{4\pi}\int_{-\infty}^\infty d\tau \int_0^{2\pi}
d\sigma\left(
\partial_\tau X^2+\partial_\sigma X^2+\partial_\tau
Y^2+\partial_\sigma Y^2\right)
\end{equation}
and the boundary condition is now the two equations
\begin{subequations}
\label{generallabel}
\begin{eqnarray}
\left[ -\frac{1} {2\pi} \partial_\tau X(0,\sigma)+i\frac{g}{2}
e^{iX(0,\sigma)}-i
\frac{\bar g}{2} e^{-iX(0,\sigma)}\right]\BDket &=&0 \label{bbc100}\\
Y(0,\sigma)\BDket &=&0. \label{dbc100}
\end{eqnarray}
\end{subequations}
The theory for
$Y$ is decoupled and is exactly solvable.  Its contribution to
quantities such as the partition function is easily identified,
factored out and discarded to return to the theory with $X$ alone.
The presence of $Y$ is needed to make the mapping to fermion
variables described below. It allows us to form the variables
\begin{equation}\phi_1 = \frac{1}{\sqrt{2}}\left( X+Y\right)~~,~~
\phi_2=\frac{1}{\sqrt{2}}\left(X-Y\right)\label{variables}\end{equation}
with which the mapping to fermions is defined:
\begin{subequations}
\label{generallabel}
\begin{eqnarray}\label{fermionization1}
\psi_{1L}(z)&=&\zeta_{1L}:e^{-\sqrt{2}i\phi_{1L}(z)}:
~~,~~\psi_{1L}^\dagger(z)=:e^{\sqrt{2}i\phi_{1L}(z)}: \zeta_{1L}^\dagger\\
\psi_{2L}(z)&=&\zeta_{2L}:e^{\sqrt{2}i\phi_{2L}(z)}: ~~,~~
\psi_{2L}^\dagger(z)=:e^{-\sqrt{2}i\phi_{2L}(z)}:\zeta_{2L}^\dagger\\
 \psi_{1R}(\bar z)&=&\zeta_{1R}:e^{\sqrt{2}i\phi_{1R}(\bar z)}: ~~,~~
  \psi_{1R}^\dagger(\bar z)=:e^{-\sqrt{2}i\phi_{1R}(\bar z)}:\zeta_{1R}^\dagger \\
\psi_{2R}(\bar z)&=& \zeta_{2R} :e^{-\sqrt{2}i\phi_{2R}(\bar
z)}:~~,~~ \psi_{2R}^\dagger(\bar z)= :e^{\sqrt{2}i\phi_{2R}(\bar
z)}:\zeta_{2R}^\dagger\label{fermionization2}
\end{eqnarray}
\end{subequations}
where $\zeta_{aL/R}$ are co-cycles. Co-cycles must be introduced
in order to make the fermion fields anti-commute with each other.
An explicit representation of the co-cycles constructed from the
zero modes of the bosons,
$\sqrt{2}\phi_{aL/R}=\varphi_{aL/R}-i\pi_{aL/R}\ln z+\ldots$ is
given in ref.~\cite{Lee:2005ge},
\begin{eqnarray}
\zeta_{1L}=\zeta_{1R}=\exp\left(-i\frac{\pi}{2}\left(\pi_{1L}+\pi_{1R}+2\pi_{2L}+2\pi_{2R}\right)\right)
 ,~   \zeta_{2L}=\zeta_{2R}=
  \exp\left(-i\frac{\pi}{2}\left(\pi_{2L}+\pi_{2R}\right)\right)
  \label{cocycle}
\end{eqnarray}
The fermion bilinear operators are
\begin{subequations}
\label{generallabel}
\begin{eqnarray}
:\psi^\dagger_{1L}(z)\psi_{1L}(z):= i\sqrt{2}\ditau\phi_{1L}(z)
~~&,&~~ :\psi^\dagger_{2L}(z)\psi_{2L}(z): =
-i\sqrt{2}\ditau\phi_{2L}(z)\label{bilinears1}\\
:\psi^\dagger_{1R}(\bar z)\psi_{1R}(\bar z): =
-i\sqrt{2}\ditau\phi_{1R}(\bar z)  ~~&,&~~ :\psi^\dagger_{2R}(\bar
z)\psi_{2R}(\bar z): = i\sqrt{2}\ditau\phi_{2R}(\bar
z)\label{bilinears2}. \end{eqnarray}
\end{subequations}
The fermion action on the infinite cylinder is
$$
S=\frac{i}{2\pi}\int_{-\infty}^\infty d\tau \int_0^{2\pi}d\sigma
\left( \psi_{aL}^\dagger(\partial_\tau+i\partial_\sigma)\psi_{aL} +
\psi_{aR}^\dagger(\partial_\tau-i\partial_\sigma)\psi_{aR}\right)
$$
%
The fermion field theory has an explicit $SU(2)_L\times SU(2)_R$
chiral symmetry which, because the currents are holomorphic, extends
to a linear representation of two copies of the level-1 SU(2)
Kac-Moody algebra generated by
\begin{eqnarray}
J^a_L(z)= \frac{1}{2}:\psi^{\dagger}_{L}(z)\sigma^a\psi_{L}(z):
~~~~~~\sigma^a={\rm ~Pauli~matrices}\label{km} \\
{J'}^+_L(z)=\psi^{\dagger}_{L1}(z)\psi^{\dagger}_{L2}(z) ~~,~~
{J'}^-_L(z) = \psi_{L2}(z)\psi_{L1}(z) ~~,~~
{J'}^3_L(z)=\frac{1}{2}:\psi_L^{\dagger}(z)\psi_L(z):
\label{kmprime}\end{eqnarray} The current algebras generated by
$J^a_L(z)$ and ${J'}^a_L(z)$ are identical to the well-known bosonic
current algebras for the $X$- and $Y$-bosons, respectively. There
are also anti-holomorphic currents, $J^a_R(\bar z)$ and
${J'}^a_R(\bar z)$.

The mapping of fermion states to boson states is not generally 1-1.
When the fermions are quantized, the eigenvalues of the fermion
number operators, which are identified with boson momenta, are
spaced by integers.  This suggests that the fermions correspond to
compact bosons.  Indeed, it was shown in ref.~\cite{Lee:2005ge}
that, to obtain the states of the bosons compactified at the
self-dual radius (\ref{radius}) and when $Y$ is also identified in
the same way, $Y\sim Y+2\pi$, we must consider fermions in two
sectors, the Ramond (R) sector where they all have periodic boundary
conditions on the cylinder and the Neveu-Schwarz (NS) sector where
all are anti-periodic. Then we must project onto states with even
total fermion number. The latter is the analog of the GSO projection
of the worldsheet fermions in the Neveu-Schwarz-Ramond formulation
of the superstring which obtains the type 0A and 0B theories.
Indeed, it is known to be the unique projection which produces a
modular invariant partition function in D=1 \cite{Seiberg:1986by}.

If the bosons are not compact, we shall first treat them as
compact and solve for their partition function which will turn out
to depend on the Wilson line variable $A$. Then, in accord with
our discussion in Sect.~1.1, we integrate the partition function
over allowed values of $A$.

\subsection{Boundary states in fermion variables}

We shall write the boundary conditions in eqs.~(\ref{bbc100}) and
(\ref{dbc100}) in fermion variables.  The first term is $ \ditau
X(0,\sigma) = \frac{1}{2i}\left[ :\psi^{\dagger}_L(0,\sigma)
\sigma^3\psi_L(0,\sigma): - :\psi^{\dagger}_R(0,\sigma)
\sigma^3\psi_R(0,\sigma):\right] $. Moreover, we can also use the
Dirichlet condition for $Y$ (\ref{dbc100}) in the boundary
interaction,
\begin{eqnarray}
e^{iX(0,\sigma)}\BDket  &=& e^{i(X(0,\sigma)+Y(0,\sigma))}\BDket
\nonumber\\
&=& e^{\sqrt{2}i \phi_{1L}(0,\sigma)}e^{
\sqrt{2}i\phi_{1R}(0,\sigma)}\BDket \nonumber \\
&=& Z^2 :e^{\sqrt{2}i \phi_{1L}(0,\sigma)}::e^{
\sqrt{2}i\phi_{1R}(0,\sigma)}:\BDket \\
&=& Z^2
\psi^{\dagger}_{1L}(0,\sigma)\zeta_{1L}\zeta^{\dagger}_{1R}\psi_{1R}(0,\sigma)\BDket
\nonumber\\
&=&  \tfrac{1}{2}Z^2
\psi^{\dagger}_L(0,\sigma)\left(1+\sigma^3\right)\psi_R(0,\sigma).
\BDket \nonumber
\end{eqnarray}
Here, when we use the explicit representation of the co-cycles
(\ref{cocycle}), they cancel. $Z$ is an infinite constant which
accounts for the difference between the operators and normal
ordered operators. We shall absorb $Z^2$ by re-defining the
coupling constant $g$. This provides the correct power of the
cutoff to make $g$ a marginal coupling and is the only
renormalization that is required in this model.

In addition, the last term in the boundary interaction
is\footnote{Note that we could have also fermionized $e^{-i(X+Y)}$ to
get the operator
$\psi^{\dagger}_R\frac{1}{2}\left(1+\sigma^3\right)\psi_L$.
Consistency requires that this operator, when acting on the boundary
state, is equivalent to the one in (\ref{gbarferm}). It is possible to
use the solution for $|BD>$ that we shall find to demonstrate that
this is indeed the case.}
\begin{eqnarray}
e^{-iX(0,\sigma)}\BDket  =  e^{-i(X(0,\sigma)-Y(0,\sigma))}\BDket
 =  \tfrac{1}{2}Z^2
\psi^{\dagger}_L(0,\sigma)\left(1-\sigma^3\right)\psi_R(0,\sigma)\BDket
\label{gbarferm}
\end{eqnarray}
    The boundary state condition (\ref{bbc100}) becomes
\begin{eqnarray}\label{fermbsc} \left[:
\psi^{\dagger}_{L} \sigma^3 \psi_{L}: - :\psi_{R}^{\dagger} \sigma^3
\psi_{R}: +\pi g\psi^{\dagger}_L\left(1+\sigma^3\right)\psi_R
  - \pi\bar g\psi^{\dagger}_L\left(1-\sigma^3\right)\psi_R \right]\BDket & = & 0
\end{eqnarray}

Our task is now to find a state in the fermion theory which solves
(\ref{fermbsc}) and retains the Dirichlet boundary condition for the
$Y$-boson. It is known that
some boundary states are related to each other by global SU(2)
rotations of the left-handed fields. For example, the DD and ND
boundary conditions
\begin{subequations}
\label{generallabel}
\begin{eqnarray}
\left(\psi_{R}(0,\sigma)-\psi_{L}(0,\sigma)\right)|DD>=0&,&
\left(\psi^\dagger_{R}(0,\sigma)
+\psi^{\dagger}_{L}(0,\sigma)\right)|DD>=0
\label{DDferm}\\
\left(\psi_R(0,\sigma)+i\sigma^1\psi_L(0,\sigma)\right)|ND>=0
&,& \left(\psi^{\dagger}_R(0,\sigma)
+\psi^{\dagger}_L(0,\sigma)i\sigma^1\right)|ND>=0 \label{NDferm}
\end{eqnarray}
\end{subequations}
can be found directly from the boson-fermion mapping in
eqs.~(\ref{fermionization1})-(\ref{fermionization2}), the N and D
conditions for X and Y in (\ref{strongform}) and the co-cycles
(\ref{cocycle}). Solving the simple equations (\ref{DDferm}) and
(\ref{NDferm}) (see ref.~\cite{Lee:2005ge}) obtains the boundary
states in fermion variables,  These are identical to the states
which would be obtained by the appropriate direct product of the N
and D boundary states which we have written in boson variables in
eqs.~(\ref{bstates1})-(\ref{bstates2}).

The DD and ND boundary conditions in (\ref{DDferm}) and
(\ref{NDferm}) are related by a global $SU(2)$ rotation. With the
appropriate choice of phases, the boundary states must also be
related by the same transformation,
\begin{equation} |DD>=e^{-i\pi J^1_L}|ND>\label{DD-ND}
\end{equation} where $J^a_L=\oint\frac{
d\sigma}{2\pi} J^a_L(0,\sigma)$ are the generators of the $SU(2)$
algebra. It was argued in ref.~\cite{Callan:1994ub} that $|B>$ can
also be obtained from $|N>$ by a global rotation.  To examine this
possibility in the present context, we begin with the ansatz
\begin{equation}\label{ansatz}
|BD> = e^{-i\theta\cdot J_L }|ND> \end{equation} We shall find
that for the most general case, we shall require complex values of
$\theta^a$, and hence a transformation in $SL(2,C)$. Since $J^a_L$
operates on the $X$-boson only, the transformation of the $|ND>$
state in (\ref{ansatz}) does not upset the Dirichlet boundary
condition for the $Y$-boson. The transformation operates on the
fermions by
\begin{eqnarray}\label{angles}
 e^{-i\theta\cdot J_L }\psi_L
 e^{i\theta\cdot J_L } = U\psi_L
~,~e^{-i\theta\cdot J_L }\psi^\dagger_L
  e^{i\theta\cdot J_L } = \psi^\dagger_LU^{-1}
~,~U=e^{ i\theta\cdot\sigma/2}
 \end{eqnarray}
 The fact that the transformation operates on the $X$-boson only and
 leaves the $Y$-boson unchanged is encoded in
 $\det[U]=1$.
 The postulate (\ref{ansatz}) is equivalent to the statement that, like
 DD and ND boundary conditions which are presented in eqs.~(\ref{DDferm}) and (\ref{NDferm})
 as the usual gluing conditions of boundary conformal field theory,
the boundary state $|BD>$ also obeys a gluing condition,
 \begin{equation}\label{gluing}
 \left( \psi_R(0,\sigma) +i\sigma^1 U\psi_L(0,\sigma)\right)|BD>=0
 ~~,~~
 \left(\psi_R^\dagger(0,\sigma) +\psi_L^\dagger(0,\sigma)U^{-1}i\sigma^1\right)|BD>=0
 \end{equation}
 This class of boundary conditions automatically imply that $|BD>$ is a conformal
 boundary state, since the energy momentum tensor is an $SU(2)\times SU(2)$
 invariant quadratic in
 fermion operators, $$ \left( L_n - \tilde L_{-n}\right)|BD>=0$$

It is easy to see that the boundary condition (\ref{gluing})
solves eq.~(\ref{fermbsc}) when $U$ is
 \EQA{\label{solution}
    U =  \twobytwo{ e^{-2\pi i A}\sqrt{1-\pi^2 g\bar g} & -i\pi g}{-i\pi \gbar &
     e^{2\pi i A}\sqrt{1-\pi^2 g\bar g} }
}

The constant $A$ is not determined by eq.~(\ref{fermbsc}). However,
certain partition functions constructed using the boundary state
depend on it.   We will argue later that it is indeed the
topological parameter of the action (\ref{bosonaction}) and we have
anticipated this by the notation. $U\in$SL(2,C) is a complex matrix
with unit determinant. It is unitary (and the angles in
(\ref{angles}) real -- this is needed for unitarity of the conformal
field theory) when $A$ is real and when $\bar g=g^*$ and
$|g|<1/\pi$. When the coupling exceeds a critical strength,
$|g|>1/\pi$, the matrix is not unitary.

Note, by comparing with (\ref{DD-ND}), that the state $|BD>$ becomes
the state $|DD>$ when $U=i\sigma^1$.  By inspecting (\ref{solution})
we see that this happens at the critical coupling strength, when
$g=-1/\pi=\bar g$. This can be generalized to $g=-e^{i\phi}/\pi$ and
$\bar g = -e^{-i\phi}/\pi$. Then, the phase $\phi$ corresponds to
the D-brane position on the circle, or if the boson is not compact,
the positions of an infinite array of equally spaced D-branes.

We can now construct the boundary state explicitly using the
fermions. The quantization of fermions uses the mode expansion
\begin{subequations}
\label{generallabel}
\begin{eqnarray}\psi_{La}(z)=\sum_r \psi_{a,r} z^{-r} ~~&,&~~\psi_{Ra}(\bar
z)=\sum_r\tilde\psi_{a,r}{\bar z}^{-r}\\
\psi_{La}^\dagger(z)=\sum_r \psi_{a,r}^\dagger z^{-r}
~~&,&~~\psi_{Ra}^\dagger(\bar z)=\sum_r\tilde\psi_{a,r}^\dagger{\bar
z}^{-r}
\end{eqnarray}
\end{subequations}
where $r$ are half-odd integers in the NS sector and integers in the
R sector.  Non-vanishing anti-commutators are
\begin{eqnarray} \left\{
\psi_{a,r},\psi^\dagger_{b,s}\right\}=\delta_{ab}\delta_{rs} ~~,~~
\left\{
\tilde\psi_{a,r},\tilde\psi^\dagger_{b,s}\right\}=\delta_{ab}\delta_{rs}
\end{eqnarray}
The vacuum states are annihilated by all positively moded operators
and care in handling the zero modes in the R sector is
required.\footnote{ We define the R sector state $|---->$ as the one
  which is annihilated by all positively moded oscillators and the
  zero mode operators
  $\left(\psi_{1,0},\tilde\psi_{1,0},\psi_{2,0},\tilde\psi_{2,0}\right)$,
  respectively.  Then $\psi_{a,0}^{\dagger}$ or
  $\tilde\psi_{a,0}^\dagger$ operating on this state flips a - to a +
  in the appropriate position. The phase is defined as having a plus sign when
the creation operators are ordered as $\left(\psi_{1,0}^\dagger,
\tilde\psi_{1,0}^\dagger,\psi_{2,0}^\dagger,\tilde\psi_{2,0}^\dagger\right)$,
In total there are 16 degenerate ground
  states of the fermion system. For each species of fermion, the fermion
number of $|\pm>$ is $\pm\tfrac{1}{2}$.} In the N-S sector the
$|BD>$ state is\footnote{Some of the
  phase conventions differ from ref.~\cite{Lee:2005ge}.}
\begin{equation}\label{nsboundarystate}
|BD>_{NS}=2^{-\tfrac{1}{2}}\prod_{r=\tfrac{1}{2}}^\infty
\exp\left[ \psi_{-r}^\dagger U^{-1}i\sigma^1\tilde\psi_{-r} -
\tilde\psi^{\dagger}_{-r}i\sigma^1U\psi_{-r}\right]~|0>
\end{equation}
and in the R-sector it is
\begin{equation}\label{rboundarystate}
|BD>_R=2^{-\tfrac{1}{2}}\prod_{n=1}^\infty  \exp\left[
\psi_{-n}^\dagger U^{-1}i\sigma^1\tilde\psi_{-n} -
\tilde\psi^{\dagger}_{-n}i\sigma^1U\psi_{-n}\right]\exp\left[\psi_0^\dagger
U^{-1}i\sigma^1\tilde\psi_0\right]~|-+-+>
\end{equation}
The factor of $2^{-\tfrac{1}{2}}$ has the same origin as the factor
of $2^{-\tfrac{1}{4}}$ in the boson boundary states
(\ref{bstates1})-(\ref{bstates2}). The Hamiltonian in the NS and R
sectors are
\begin{subequations}
\label{generallabel}
\begin{eqnarray}
\hat H_{NS}&=&\sum_{r=\frac{1}{2}}^\infty r\left(\psi_{-r}^\dagger\psi_r +
\psi_{-r}\psi_r^\dagger
+ \tilde\psi_{-r}^\dagger \tilde
\psi_r+ \tilde\psi_{-r}\tilde\psi_r^\dagger \right)-\frac{1}{6} \\ \hat
H_{R}&=&\sum_{n=1}^\infty n\left(\psi^\dagger_{-n}\psi_n+\psi_{-n}\psi^\dagger_n
+\tilde\psi^\dagger_{-n}\tilde
\psi_n + \tilde\psi_{-n}\tilde\psi^\dagger_n\right)+\frac{1}{3}
\end{eqnarray}
\end{subequations}
respectively. In the NS sector,
\begin{eqnarray}
e^{-\alpha \hat H}  |BD>=
2^{-\tfrac{1}{2}}e^{\tfrac{\alpha}{6}}\prod_{r=\tfrac{1}{2}}^\infty
\exp\left[ e^{-2\alpha r}\psi_{-r}^\dagger
U^{-1}i\sigma^1\tilde\psi_{-r} - e^{-2\alpha r}
\tilde\psi^{\dagger}_{-r}i\sigma^1U\psi_{-r}\right]~|0>.
\end{eqnarray}

As a concrete example, we will first compute the matrix element of
this state with the $|DD>$ boundary state. The $|DD>$ state is
gotten from $|BD>$ simply by replacing $U$ by $i\sigma^1 $, to get
\begin{equation}
{}_{NS}<DD|~=2^{-\tfrac{1}{2}}<0| \prod_{r=\tfrac{1}{2}}^\infty
\exp\left[- \tilde\psi^{\dagger}_{r}\psi_{r}+\psi_{r}^\dagger
\tilde\psi_{r} \right].
\end{equation}
(To take the conjugate, we have to know
that the conjugate of
$\psi_{-r}$ is $\psi_r^\dagger$, etc.)  The partition function of
interest is
\begin{eqnarray}
Z &=& <DD|~e^{-\alpha\hat H}|BD> \nonumber\\
&=&\tfrac{1}{2}e^{\tfrac{\alpha}{6}}
<0|\prod_{r=\tfrac{1}{2}}^\infty e^{-\psi_r^\dagger\tilde
\psi_r+\psi_r^\dagger\tilde\psi_r} e^{e^{-2\alpha
r}\psi^\dagger_{-r}U^{-1}i\sigma^1 \tilde\psi_{-r} - e^{-2\alpha
r}\tilde\psi^\dagger_{-r} i\sigma^1U\psi_{-r}}|0>.
\end{eqnarray}
It is easy to find this matrix element by computing it for one
value of $r$ at a time.  To simplify further, in Section 3, we shall show that it depends only on
the eigenvalues of $-i\sigma^1U$ which, since the determinant is
one, are a complex number $\zeta $ and its inverse $1/\zeta$ where
\begin{equation}\zeta=\frac{ \pi(g+\bar g) } {2}
+i\sqrt{ 1-\left(\frac{ \pi(g+\bar g) }{ 2 }
\right)^2}\label{zeta}\end{equation} Note that these eigenvalues do
not depend on the  parameter $A$ in (\ref{solution}). The NS sector
partition function is \footnote{We use the Jacobi triple product
formula
$$
\sum_{n\in{\cal Z}}z^n q^{n^2} = \prod_{n=0}^\infty
\left(1-q^{2n+2}\right)\left(1+zq^{2n+1}\right)\left(1+z^{-1}q^{2n+1}\right)
$$}
\begin{eqnarray} \left.
<DD|~e^{-\alpha\hat
H}|BD>\right|_{NS} &=& \tfrac{1}{2}e^{\tfrac{\alpha}{6}}\prod_{r=\frac{1}{2}}^\infty
\left(1+\zeta e^{-2\alpha r}\right)^2\left(1+\zeta^{-1}e^{-2\alpha
r}\right)^2 \nonumber \\
&=& \tfrac{1}{2}e^{\frac{\alpha}{6}}\left[ \sum_n \zeta^n e^{-\alpha
n^2} \prod_{k=1}^\infty \frac{1}{\left( 1-e^{-2\alpha
k}\right)}\right]^2
\end{eqnarray}
Similarly, in the R sector, the partition function is
\begin{eqnarray} \left. <DD|~e^{-\alpha\hat H}|BD>\right|_{R}
&=& \tfrac{1}{2}e^{-\frac{\alpha}{3}}
\prod_{n=1}^\infty \left(1+\zeta e^{-2\alpha
n}\right)^2\left(1+\zeta^{-1}e^{-2\alpha n}\right)^2\cdot
(1+\zeta)(1+\zeta^{-1}) \nonumber
\\
&=& \tfrac{1}{2}e^{\frac{\alpha}{6}}\left[ \sum_n
\zeta^{\left(n+\tfrac{1}{2}\right)}
e^{-\alpha\left(n+\frac{1}{2}\right)^2} \prod_{k=1}^\infty
\frac{1}{\left( 1-e^{-2\alpha k}\right)}\right]^2
\end{eqnarray}
In the first line above, the product on the right-hand-side is the
contribution of non-zero modes and the last factors are the
contribution of the zero modes.

We must still remember to project onto the states whose total
fermion number is even.  The boundary state $|ND>$ has even fermion
number and neither the operation of the Hamiltonian nor the SL(2,C)
transformation changes that.  Therefore, no projection is needed.
The full partition function is obtained by combining the NS and R
sectors,
\begin{eqnarray} Z &=& \left.
<DD|~e^{-\alpha\hat H}|BD>\right|_{NS}+\left. <DD|~e^{-\alpha
H}|BD>\right|_{R} \nonumber \\
&=& \tfrac{1}{2}e^{\frac{\alpha}{6}} \left( \prod_{k=1}^\infty
\frac{1}{\left(1-e^{-2\alpha k}\right)} \right)^2 \left(
 \left[ \sum_n \zeta^n e^{-\alpha n^2} \right]^2+\left[
\sum_n \zeta^{\left(n+\tfrac{1}{2}\right)} e^{-\alpha
\left(n+\frac{1}{2}\right)^2}\right]^2 \right) \nonumber \\
&=&\tfrac{1}{2}e^{\frac{\alpha}{6}} \left( \prod_{k=1}^\infty
\frac{1}{\left(1-e^{-2\alpha k}\right)} \right)^2
 \sum_{m,n} \zeta^{m-n}\left( e^{-\alpha (m^2+n^2)}
 +e^{-\alpha\left(m+\frac{1}{2}\right)^2-\alpha
 \left(n+\frac{1}{2}\right)^2} \right) \\
&=&\tfrac{1}{2}e^{\frac{\alpha}{6}} \left( \prod_{k=1}^\infty
\frac{1}{\left(1-e^{-2\alpha k}\right)} \right)^2
 \sum_{m,n} e^{-\frac{1}{2}\alpha\left(m-n\right)^2}
 \zeta^{m-n}\left( e^{-\frac{1}{2}\alpha\left(m+n\right)^2}
 +e^{-\frac{1}{2}\alpha\left(m+n+1\right)^2}\right)\nonumber
\end{eqnarray}
When m and n run over all of the integers, m+n, m-n run over all
pairs of integers where both are  even or both are odd and m+n+1,m-n
run over all pairs of integers where one is even and the other is
odd. Thus, the last expression is
\begin{eqnarray} Z=
\left(\tfrac{1}{\sqrt{2}}e^{\frac{\alpha}{12}}\sum_n
e^{-\frac{1}{2}\alpha n^2} \prod_{k=1}^\infty
\frac{1}{\left(1-e^{-2\alpha k}\right)} \right)\cdot
\left(\tfrac{1}{\sqrt{2}}e^{\frac{\alpha}{12}}\sum_n \zeta^n
e^{-\frac{1}{2}\alpha n^2} \prod_{k=1}^\infty
\frac{1}{\left(1-e^{-2\alpha k}\right)} \right)
\end{eqnarray}
The first factor on the right-hand-side is the partition function of
the $Y$-boson which we discard.  The remaining partition function is
\begin{equation}
<D|~ e^{-\alpha\hat H}|B>=\tfrac{1}{
\sqrt{2}}e^{\frac{\alpha}{12}}\sum_n \zeta^n
e^{-\frac{1}{2}\alpha n^2} \prod_{k=1}^\infty
\frac{1}{\left(1-e^{-2\alpha k}\right)}
\end{equation}
Now, we can recover the partition function of the field theory on
a strip if, according to (\ref{bspart}), we put
$\alpha=2\pi^2/\beta$. Then, a modular transformation, Poisson
resummation
\footnote{
The Dedekind eta-function which is defined by
$$
\eta(\tau)\equiv e^{2\pi i\tau/24}\prod_{n=1}^\infty
\left(1-e^{2\pi i\tau n}\right)
$$
 has the modular transformations
 $$
\eta(\tau+1)=e^{2\pi i/24}\eta(\tau) ~~,~~
\eta(-1/\tau)=(-i\tau)^{\tfrac{1}{2}}\eta(\tau)$$  The Poisson
re-summation is
$$
\sum_{p\in{\cal Z}}
e^{-ap^2+ibp}=
\sqrt{\tfrac{\pi}{a}}
\sum_{n\in{\cal Z}}
e^{-\tfrac{\pi^2}{a} ( n+b/2\pi)^2}~~.
$$
}
and using (\ref{zeta}) give
\begin{equation}
Z_{DB}[\beta]= e^{\frac{\beta}{24}}\sum_n   e^{-\beta
\left(n+\delta\right)^2} \prod_{k=1}^\infty
\frac{1}{\left(1-e^{-\beta k}\right)} ~~,~~\delta=
\frac{1}{2\pi}\cos^{-1} \frac{\pi}{2}(g+\bar g)
\label{partitionfunction}
\end{equation}
This coincides with the results quoted in \cite{Callan:1994ub} and
\cite{Polchinski:1994my} if we make a certain re-definition of the
coupling constants $(g,\bar g)$.  We shall discuss this
re-definition in the next Section.

\section{Partition functions and spectrum}

 In this section, we shall summarize some comments about what we
 have done so far.

\begin{enumerate}
\item{}{\bf Wilson line:} The source of the parameter $A$ in
(\ref{solution}) is clear.  The boundary condition (\ref{bsgbs})
does not fix the dependence of $|B>$ on the combination of zero
modes $x_L-x_R$. It is easy to see that, if $\theta^a$ can be
chosen so that $|BD>=e^{-i\theta\cdot J_L}|ND>$ satisfies the
boundary state equation, then
\begin{equation}
|BD>_a = e^{-ia J^3_L}
e^{-i\theta\cdot J_L} e^{-ia J^3_L} |ND>
\end{equation}
also satisfies the same equation.   This replaces $2\pi A$ by
$2\pi A +a$  in the solution (\ref{solution}). To see that this
amounts to a translation of $x_L-x_R$, we simply observe that we
can use the Neumann boundary condition to write this state as
\begin{equation}
|BD>_a =e^{-ia J^3_L}e^{-i\theta\cdot J_L} e^{-ia J_R^3} |ND>
= e^{-ia( J^3_L+J_R^3)}e^{-i\theta\cdot J_L}  |ND>
\end{equation}
and that $J^3_L+J_R^3$ generates a translation of $x_L-x_R$,
leaving $x_L+x_R$, $y_L$ and $y_R$ and all other boson modes
unchanged.  If we write $ |BD>_A =e^{-2\pi iA (J^3_L +
J^3_R)}|BD>_0$, we see that, indeed
\begin{equation}
e^{-2\pi iA(J_L^3+J_R^3)}=e^{A\int d\sigma\left(\partial_\tau X_L
- \partial_\tau X_R\right)}~=~e^{ -iA\oint d\sigma
\partial_\sigma(X_L+X_R)}
\end{equation}
inserts the operator $e^{ -iA\oint d\sigma
\partial_\sigma(X_L+X_R)}
$ at the boundary and produces the topological term in the action
(\ref{bosonaction}).

\item{} {\bf Partition function with general boundary states:} In
  ref.~\cite{Gaberdiel:2001xm,Gaberdiel:2001zq} it has been argued that the general
  conformal boundary state of a system which satisfies the consistency
  conditions of boundary conformal field theory with a single free
  boson compactified at the self-dual radius is the one obtained by
  the SL(2,C) transformation $ |\theta > = e^{- i\theta\cdot J_L}|N>
  $.\footnote{For a review of boundary conformal field
  theory in similar contexts, see \cite{Schomerus:2005aq} and one
  which discusses the particular boundary states of interest here,
  see \cite{Gaberdiel:2005sk}.}
  It is generally specified by three complex parameters, $\theta^a$.
  When $\theta^a$ are complex, the conjugate of this state is defined
  as $<\theta|\equiv <N|e^{i\theta\cdot J_L}$. The partition function
  between two such states is
\begin{equation}\label{general}
<\theta_1 |~e^{-\tfrac{2\pi^2}{\beta} \hat H}|\theta_2 > =
e^{\frac{\beta}{24}}\sum_{n\in {\cal Z}} e^{- \beta (n+\delta)^2}
\prod_{k=1}^\infty \frac{1}{\left(1-e^{-\beta k}\right)}
\end{equation}
where $(e^{2\pi i\delta},e^{-2\pi i\delta})$ are eigenvalues of $
U_2U_{1}^{-1} $ and $U_i=e^{i\theta_i\cdot \sigma/2}$.  The result
is quoted in eqs.~(\ref{theanswer}) and (\ref{theanswer2}).
\begin{itemize}
Some interesting special cases can be found by taking limits of
eqs.~(\ref{theanswer}) and (\ref{theanswer2}).  \item{}The case
where one of the boundaries has a Dirichlet condition,
$U_1=i\sigma^1$, agrees with (\ref{partitionfunction}).
 \item{}When one of the boundaries has a Neumann boundary condition,
$U_1=1$,
$Z_{N\theta}[\beta]=~_{A_1}<N|~e^{-\tfrac{2\pi^2}{\beta}\hat
    H}|B>_{A_2}$, is given by (\ref{general}) with
  \begin{equation}\label{nbs}\delta=\frac{1}{2\pi}\cos^{-1}\left(\sqrt{1-\pi^2
      g\bar g}\cos2\pi (A_1-A_2)\right) \end{equation}
      \item{}For the half-brane, $g\neq 0$ and $\bar
  g=0$, the partition function is given by eq.~(\ref{theanswer})
  with $\delta = \left(A_1-A_2\right)$. This agrees with the known
  result that the open-string partition function for the half-brane is
  independent of $g$.
  See ref.~\cite{Larsen:2002wc,Constable:2003rc,Gaberdiel:2004na} for a
  discussion of this point and a computation using the bosonic version
  of the boundary state.

\end{itemize}

\item{}{\bf Partition function depends only on eigenvalues:} It is
easy to see why the partition function (\ref{general}) should only
depend on the eigenvalues of the matrix $U_2U_1^{-1}$. The
boundary state $|ND>$ is strictly invariant under the simultaneous
left- and right- SL(2,C) rotation $ |ND> = e^{i\hat\theta\cdot
J_R} e^{-i{\theta}\cdot{J_L}} |ND> $ where
$(\hat\theta_1,\hat\theta_2,\hat\theta_3)=(\theta_1,-\theta_2,-\theta_3)$
and $\theta^a$ are real numbers. The reason for this is the fact
that, for example in the NS sector, using fermi statistics, the
boundary state can be written as

\begin{equation}\label{bs}
|ND>=\prod_{r=\tfrac{1}{2}}^\infty\exp\left( \psi_{-r}^\dagger
i\sigma^1\tilde\psi_{-r}-
\tilde\psi_{-r}^{\dagger}i\sigma^1\psi_{-r}\right)|0>
\end{equation}
It is easy to see that, for example $ \left[J^2_L-J^2_R,
\psi_{-r}^\dagger i\sigma^1\tilde\psi_{-r}-
\tilde\psi_{-r}^{\dagger}i\sigma^1\psi_{-r}\right]=0 $ Another way
to see this is to note that the $ND$ boundary condition for the
fermions (\ref{NDferm}), which is easily seen to be enforced when
operating on the state in (\ref{bs}), is preserved by the
simultaneous rotations
$$
\psi_r \to e^{i\theta\cdot\sigma/2}\psi_r~~,~~\psi_r^\dagger \to
\psi^\dagger_r e^{-i\theta\cdot\sigma/2}~~,~~\tilde\psi_{-r}\to
e^{i\hat\theta\cdot\sigma/2}\tilde\psi_{-r}
~~\tilde\psi_{-r}^\dagger\to \tilde\psi_{-r}^\dagger
e^{-i\hat\theta\cdot\sigma/2}  $$ since
$\sigma^1e^{i\theta\cdot\sigma/2}\sigma^1 =
e^{i\hat\theta\cdot\sigma/2}$.

Then,
$$<ND|~...~|ND>~=~
<ND|e^{-i\hat\theta\cdot J_R}e^{i\theta\cdot J_L}~...~
e^{i\hat\theta\cdot J_R}e^{-i\theta\cdot J_L}|ND>
$$
 Since $J^a_R$ commutes with $H$ and $J_L^a$, the contents
 of $``...''$ in the above equation,
we see that
$$<ND|~...~|ND>~=~
<ND|e^{i\theta\cdot J_L}~...~ e^{-i\theta\cdot J_L}|ND>
$$
This symmetry implies that the partition function depends only on
equivalence classes of pairs of states, with the equivalence
relation $ U_2U_{1}^{-1}\sim g U_2U_{1}^{-1}g^{-1}$.  Such a
conjugation can always be chosen so as to bring the matrix to
upper triangular form with diagonal elements the eigenvalues. The
equivalence classes are therefore parameterized by the eigenvalues
which, since the determinant of $U_2U_1^{-1}$ must be one, are of
the form $(\zeta, 1/\zeta)$ where $\zeta$ is a non-zero complex
number.

\item{}{\bf Relation to previous work}: The formal expression for
the boundary state is the exponential of the boundary potential
acting on the Neumann state as
$$ |B>= ~e^{-\int V(\phi)}|N>$$
In the fermion variables, it is straightforward to show by direct
substitution that
\begin{equation}\label{formal} |BD> =  ~ e^{-\int(
\frac{g}{2}\psi_L^\dagger\frac{1}{2}(1+\sigma^3)\psi_R +\frac{\bar
g}{2}\psi_L^\dagger\frac{1}{2}(1-\sigma^3)\psi_R )}~ |ND>
\end{equation} is a solution of the boundary state equation
(\ref{fermbsc}).
If we could then use the Neumann boundary condition directly in
the operator in the exponent to convert $\psi_R$ to $\psi_L$
there.  The terms in the exponent it would become  generators of
the global SL(2,C) symmetry and it would then be manifestly clear
that $|BD>$ is an SL(2,C) rotation of $|ND>$. However, in order to
do this, a difficulty with operator ordering must be solved. The
authors of ref.~\cite{Callan:1994ub} give an argument which shows
that all of the effects of operator ordering can be absorbed into
the renormalization of the coupling constant. What this implies is
that there exist re-defined coupling constants $ g'(g,\bar g),\bar
g'(g,\bar g)$ such that
\begin{equation}\label{formal0}  e^{-\int(
\frac{g}{2}\psi_L^\dagger\frac{1}{2}(1+\sigma^3)\psi_R +\frac{\bar
g}{2}\psi_L^\dagger\frac{1}{2}(1-\sigma^3)\psi_R )}|ND> =
e^{\chi(g',\bar g')}~e^{i\int(
\frac{g'}{2}\psi_L^\dagger\sigma^+\psi_L +\frac{\bar{ \tilde
g}}{2}\psi_L^\dagger\sigma^-\psi_L )}|ND>
\end{equation}
Our results confirm this.  We have found that, by making the
ansatz (\ref{ansatz}), the boundary state equation is indeed
solved by such a state. Comparing (\ref{formal0}) with our
solution leads to the conclusion that
\begin{equation}
U= \exp\left(i\pi g'\sigma^+ +i\pi  {\bar g}' \sigma^-\right) =
\left(
\begin{matrix} \cos \pi\sqrt{{\bar  g}' g'}
        & i\frac{g'}{\sqrt{ {\bar g}'g'} }\sin\pi\sqrt{ {\bar g}'g' }
                 \cr
 i\frac{ {\bar g}' } {\sqrt{ {\bar g}'
g}} \sin \pi \sqrt{ {\bar g}' g' }
  & \cos \pi \sqrt{ {\bar g}' g'}
 \cr
\end{matrix} \right)
\end{equation}
Comparing this with (\ref{solution}) we see that the relationship
between coupling constants is
\begin{equation}\label{constants}
\sin^2\pi\sqrt{ {\bar g}'g'} = \pi^2 \bar g g ~~,~~ \frac{g'}{{\bar
g}'} = \frac{g}{\bar g}
\end{equation}
As we would expect, since in the special case when either $g=0$ or
$\bar g=0$, operator ordering does not block the application of
the Neuman boundary condition in the exponent, the relation
becomes an identity for the other constant ($\bar{g'}=\bar g$ or $
g' = g$, respectively). (\ref{constants}) does not agree with the
transformation of coupling constants found in
ref.~\cite{Kogetsu:2004te} or ~\cite{Recknagel:1998ih} and used in
\cite{Sen:2002nu}. We attribute this to dependence on
regularization scheme and regard it as a re-parameterization of
coupling constants which should have no effect on the physical
meaning of the theory.  However, a clear understanding of how
correlators are related to the bare coupling constants could be of
value in condensed matter applications.

\item{}{\bf Band structure of the spectrum:} In
ref.~\cite{Polchinski:1994my} it was shown that, in some cases the
spectrum of a non-compact boson takes on an interesting band
structure.   To recall what the band structure is, we begin with
translation invariant theories with a free boson.  If both
boundary conditions are Neumann, for example, the boson has
momentum $p$ and energy $E=\tfrac{p^2}{2}$. In our construction,
we begin by treating it as a compact boson, whose momenta are
quantized, $\tfrac{1}{\sqrt{2}}p=n$.  The presence of the Wilson
line term in the action shifts these momenta by a constant,
$\tfrac{1}{\sqrt{2}}p=n+A$ and the energy is $E=
\left(n+A\right)^2$.  We see that the spectrum is split into
``energy bands''.  The integer $n$ labels the band and as we vary
$A$ over its range (say $-\tfrac{1}{2}$ to $+\tfrac{1}{2}$, we
sweep through the energies in a given band. Of course, for the
free boson with Neumann boundary conditions, combining the bands
just recovers the parabolic spectrum $E=\tfrac{1}{2}p^2$ with $p$
varying over the real numbers.

Now, we have observed that the addition of a boundary interaction
with discrete translation symmetry changes the boson momenta and
energies to $\tfrac{1}{\sqrt{2}}p=n+\delta(A,n)$ and $E(n,A)=
\left(n+\delta(A,n)\right)^2$, respectively. As we sweep $A$ over
its range of allowed values, $-\tfrac{1}{2}<A\leq \tfrac{1}{2}$,
$\tfrac{1}{\sqrt{2}}p=n+\delta(A,n)$ varies over a range of
momenta and $E(n,A)$ varies over the $n$'th energy band.

In eqs.~(\ref{theanswer}) and (\ref{theanswer2}) we see that
$\delta(A,n)$  does not depend on $n$, reduces to $\delta_0=A$ for
a boson with Neumann boundary conditions at both boundaries, and
otherwise is a function of the coupling constants of the boundary
sine-Gordon theory and $A$ given in (\ref{theanswer2}). The
example of the system with a Neumann boundary condition plus
boundary sine-Gordon with particular values of the coupling
constants are shown in fig.~(\ref{bands}).

When the translation invariance is broken entirely, as in the case
of a Dirichlet boundary condition, we can still think of the
position of the D-brane as a parameter analogous to $A$.
Translation invariance can be restored by integrating over this
parameter.  For example, in eq.~(\ref{partitionfunction}), if the
D-brane were placed at $X=x_0$, rather than $X=0$, the energy
shift would be replaced by
\begin{equation}
\delta= \frac{1}{2\pi}\cos^{-1} \frac{\pi}{2}\left(ge^{ix_0}+\bar
g e^{-ix_0}\right).
\end{equation}
This formula is obtained from (\ref{theanswer2}) by putting
$(g_1,\bar g_1,A_1)\to(g,\bar g,A)$ and $(g_2,\bar g_2,A_2)\to
\left( -\tfrac{1}{\pi}e^{ix_0},
-\tfrac{1}{\pi}e^{-ix_0},0\right)$.

As $x_0$ varies between $-\pi$ and $\pi$, $\delta$ sweeps across
an ``energy band''.

\end{enumerate}

\begin{figure}[htbp]
   \begin {center}
    \input{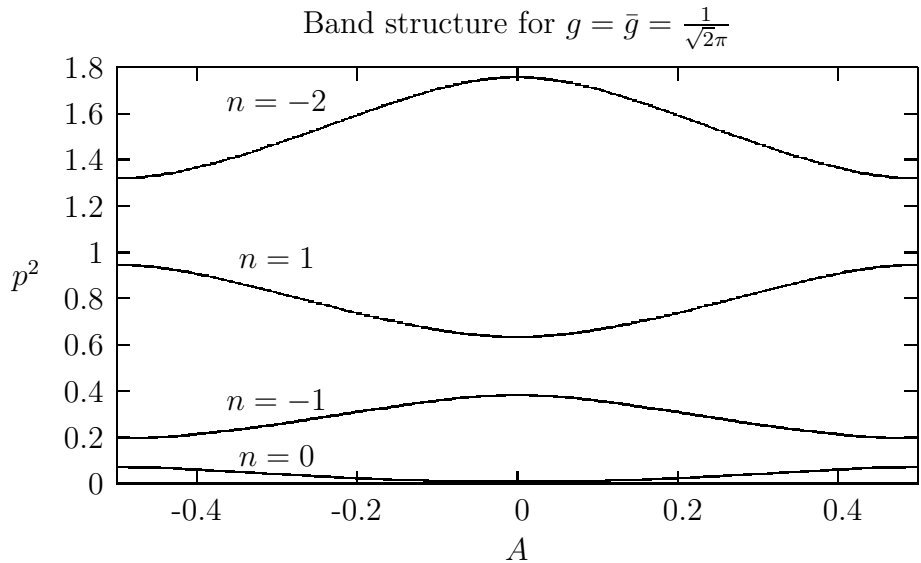}
   \end {center}
   \caption {\label{bands}The energy bands in the spectrum of a
single boson on a strip with a Neumann boundary condition at one
boundary and boundary sine-Gordon interaction on the other.}
\end{figure}

\section{Rational radii}

In the boson field theory quantized on the cylinder, the dependence
on the compactification radius of the boson resides entirely in the
allowed momentum quantum numbers.  The total momentum,
$\frac{1}{\sqrt{2}}\left(p_{XL}+p_{XR}\right)$ must come in units of
integer$/R$ whereas the dual momentum $\frac{1}{\sqrt{2}}\left(
p_{XL}-p_{XR}\right)$ must be integers$\cdot R$.  So far, we have
considered the case of self-dual radius, $R=1$.  In this section, we
will consider other radii, which are rational numbers
$R=\frac{M}{N}$ where $M$ and $N$ are co-prime integers.

It has been shown~\cite{Gaberdiel:2001xm,Gaberdiel:2001zq} that
conformal boundary states at rational radii can be found by simply
choosing that subset of the states contained in the boundary state
at the self-dual radius which have the correctly quantized momenta.
They further argue that, together with the SL(2,C) boundary states
that we have already discussed, these exhaust the set of possible
conformal boundary states of a system with one massless boson in the
bulk.

Since, in the fermion description of the theory, the momenta are
fermion numbers, this means that these states can be found by
projecting the boundary states at the self-dual radius onto sectors
with fixed fermion number. If we consider the $X$-boson with the
identification
$$ X\sim X+2\pi \frac{M}{N}
$$
where $M$ and $N$ are mutually prime, then the momentum and
wrappings are quantized as

\begin{equation}
\frac{1}{\sqrt{2}}p_X=\frac{1}{\sqrt{2}}\left(
p_{XL}+p_{XR}\right)=K\frac{N}{M} ~~,~~
\frac{1}{\sqrt{2}}w_X=\frac{1}{\sqrt{2}}\left(
p_{XL}-p_{XR}\right)=L\frac{M}{N}
\end{equation}
where $K$ and $L$ are integers. From these, we must choose states
that are already in the spectrum at the self-dual radius, i.e. where
the right-hand-sides of these equations are equal to integers. This
means that we can only find states where $K=QM$ and $L=RN$ with
$(Q,R)$ integers. That is,

\begin{equation}
 \frac{1}{\sqrt{2}}\left( p_{XL}+p_{XR}\right)=K\frac{N}{M}=QN ~~,~~
 \frac{1}{\sqrt{2}}\left( p_{XL}-p_{XR}\right)=L\frac{M}{N}=RM
\end{equation}

Now, using (\ref{variables}), (\ref{bilinears1}), (\ref{bilinears2})
and (\ref{km}), we see that, in terms of the current algebra
charges, these imply the constraints
\begin{equation}\label{const}
J^3_L - J^3_R = QN ~~,~J^3_L+J^3_R = RM
\end{equation}
where $Q$ and $R$ are any integers.  We can impose these as
constraints on the boundary state by operating with projection
operators,
\begin{equation}\label{rationalbs}
|B_{M/N},D> = \frac{1}{MN}\sum_{Q=0}^{N-1} \sum_{R=0}^{M-1} e^{ 2\pi
i\frac{Q}{N}(J^3_L-J^3_R)} e^{ 2\pi i\frac{R}{M}(J^3_L+J^3_R)} |BD>
\end{equation}
Since this projection does not involve the current algebra ${J^a}'$
in (\ref{kmprime}), it does not disturb the Dirichlet boundary
condition for the Y-boson.  We can also still take the Y-boson to be
compact with the self-dual radius.

The expression in (\ref{rationalbs}) is a sum over the SL(2,C)
boundary states.  The state in each term in the sum is of the form
\begin{eqnarray}
e^{ 2\pi i\frac{Q}{N}(J^3_L-J^3_R)} e^{ 2\pi
i\frac{R}{M}(J^3_L+J^3_R)} |BD> &=& e^{ 2\pi
i\left(\frac{Q}{N}+\frac{R}{M}\right)J^3_L}e^{-i\theta\cdot J_L} e^{
2\pi i\left(-\frac{Q}{N}+\frac{R}{M}\right)J^3_R} |ND> \nonumber\\
&=& e^{
2\pi i\left(\frac{Q}{N}+\frac{R}{M}\right)J^3_L}e^{-i\theta\cdot
J_L} e^{ 2\pi i\left(-\frac{Q}{N}+\frac{R}{M}\right)J^3_L} |ND>.
\end{eqnarray}
This effectively replaces the matrix $U$ by the transformed matrix
\begin{eqnarray}
U \to  e^{\pi i\left(\frac{Q}{N}-\frac{R}{M}\right)\sigma^3} U
e^{-\pi i\left(\frac{Q}{N}+\frac{R}{M}\right)\sigma^3}
= \left(
\begin{matrix} e^{-2\pi i(A+R/M)}\sqrt{1-\pi^2g\bar g} & -i\pi g e^{2\pi i Q/N} \cr
-i\pi\bar g e^{-2\pi i Q/N} & e^{-2\pi i (A+R/M)}\sqrt{1-\pi^2 g\bar
g} \cr
\end{matrix}\right).\label{parametermap}
\end{eqnarray}
Thus, each of the boundary states is the boundary state for a
boundary sine-Gordon theory with the replacement of parameters
\begin{equation}
g\to ge^{2\pi i Q/N}~~~,~~~ A\to A+R/M.
\end{equation} Then, a field theory of a
boson on a strip for which $|B_{M/N}D>$ in (\ref{rationalbs}) is
the correct boundary state has the partition function
\begin{equation}\label{fracpart}
Z= \frac{1}{MN}\sum_{Q=0}^{N-1}\sum_{R=0}^{M-1}\int[dX]
~e^{-S[X,M,N;Q,R]}
\end{equation}
where the Euclidean action is \begin{eqnarray}
S=\frac{1}{4\pi}\int_0^\beta d\tau \int_0^\pi d\sigma \left(
\partial_\tau X(\tau,\sigma)^2 + \partial_\sigma
X(\tau,\sigma)^2\right)+i\int_0^\beta d\tau
(A+\frac{R}{M})\partial_\tau X(\tau,0)\nonumber \\
+\int_0^\beta d\tau \left(\frac{g}{2}e^{iX(\tau,0)+2\pi i
\frac{Q}{N}} + \frac{\bar g}{2}e^{-iX(\tau,0)-2\pi i \frac{Q}{N}}
\right)
\end{eqnarray}
Note that the potential term in the action is invariant under
simultaneous translations $X\to X+2\pi\cdot$integer/N and $Q\to
Q-$integer mod N. Then, we use the prescription of Sect.~1.1 to
sum over Bloch wave-number, elongating the period from $X\to
X+2\pi\frac{1}{N}$ to $X\to X+2\pi\frac{M}{N}$.  The remaining
parameter has the identification $A\sim A+\frac{N}{M}$.

The constraint on the boundary state in (\ref{rationalbs}) is
reminiscent of a Gauss' law constraint in a gauge field theory,
with the unusual feature that it requires invariance only under
discrete subgroups of $U(1)$.  It is easy to imagine that such a
theory could be an effective field theory for a discrete gauge
theory.  Such theories are known to arise when a continuous gauge
symmetry is not fully spontaneously broken, but leaves a discrete
residual gauge group \cite{wilczek}.

\section{Disc Amplitude}

In this section, we will consider the disc correlation functions
$$ <p_{XL},p_{XR}|B>$$ These are called disc correlation functions because
they should reproduce the Euclidean path integral where the two
dimensional space-time is a disc, rather than strip, with the
boundary interaction living at the boundary of the disc, and the
bulk vertex operator $:e^{ip_{XL}X_L+ip_{XR}X_R}:$ inserted at the
center. The result is
\begin{equation}
2^{\tfrac{1}{4}}<p_{XL},p_{XR}|B> = \left\{ \begin{matrix}
[-iU_{12}]^{ \sqrt{2}p_{XL}}  & p_{XL} = p_{XR}>0  \cr
[U_{11}]^{\sqrt{2}p_{XL}} & p_{XL}=-p_{XR}>0 \cr
[U_{22}]^{-\sqrt{2}p_{XL}} & -p_{XL}=p_{XR}>0\cr
[-iU_{21}]^{-\sqrt{2}p_{XL}} & p_{XL}=p_{XR}<0\cr 0 & |p_{XL}|\neq
|p_{XR}|\end{matrix} \right. \label{momentumdisc} \end{equation}
This formula applies to both cases where either both of
$\tfrac{1}{\sqrt{2}}p_{XL}$ and $\tfrac{1}{\sqrt{2}}p_{XR}$ are
integers or both are half-odd-integers.

We can use eq.~(\ref{momentumdisc}) to compute the overlap of the
boundary state with the state which is a vacuum for all bulk boson
oscillators and which is an eigenstate of the average position of
the $X-$boson. We define the variables
$$ X_0 = \frac{1}{\sqrt{2}}\left( x_L + x_R\right) ~~,~~ \hat
X_0 = \frac{1}{\sqrt{2}}\left( x_L - x_R\right)$$ and the position
eigenstates of these operators as
\begin{equation}
|X_0,\hat
X_0>=\frac{1}{2\pi}\sum_{\pi}
e^{-i\tfrac{1}{\sqrt{2}}(p_{XL}+p_{XR})X_0 -
i\tfrac{1}{\sqrt{2}}(p_{XL}-p_{XR})\hat X_0}|p_{XL},p_{XR}>.
\end{equation}
Then, taking the matrix element and doing the geometric sums over
momenta obtains

\begin{eqnarray}
2^{\tfrac{1}{4}}<X_0,\hat X_0|B> =\tfrac{1}{2\pi} \left[
\frac{1}{1-U_{11}e^{i\hat X_0}} + \frac{1}{1-U_{22}e^{-i\hat X_0}}
+ \frac{1}{1+iU_{12}e^{iX_0}} + \frac{1}{1+iU_{21}e^{-iX_0} }
-3\right]
\end{eqnarray}

This formula can be applied to the case of a non-compact boson,
and a compact boson with self-dual and fractional radii.  The
appropriate expressions are presented in eqs.~(\ref{1})-(\ref{3}).

In order to illustrate the simplicity of the computation, we spell
it out in some detail below.  We arrive at this result be
computing the state $<p_{XL},p_{XR},p_{YL}=0,p_{YR}=0|BD>$. The
overlap with the zero momentum states for the $Y$-boson with its
Dirichlet state gives a factor of $2^{-1/4}$. Thus, the $Y$-boson
contribution is trivial and we can later cancel it by multiplying
by $2^{1/4}$.

Recall that $p_{YL}=\frac{1}{\sqrt{2}}\left(
\pi_{1L}-\pi_{2L}\right) $ and $p_{YR}=\frac{1}{\sqrt{2}}\left(
\pi_{1R}-\pi_{2R}\right) $. For this reason, we need to consider
states whose momenta obey $ \pi_{1L}=\pi_{2L}\equiv
\pi_L=\frac{1}{\sqrt{2}}p_{XL} $ and
$\pi_{1R}=\pi_{2R}\equiv\pi_R=\frac{1}{\sqrt{2}}p_{XR} $.

Also, recall that, in the NS sector, $(\pi_L,\pi_R)$ are both
integers, whereas in the R sector $(\pi_L,\pi_R)$ are both
half-odd-integers.  Both of these possibilities must be
considered. Since the boundary states have even fermion number, no
further GSO projection is needed.

The total boson momentum is related to total fermion number,
\begin{subequations}
\label{generallabel}
\begin{eqnarray}
\pi_{1L} &=& \int \tfrac{d\sigma}{2\pi}:\psi^\dagger_{1L}\psi_{1L}:
~~,~~ \pi_{1R}=-\int
\tfrac{d\sigma}{2\pi}:\psi^\dagger_{1R}\psi_{1R}: \\
\pi_{2L}&=&-\int \tfrac{d\sigma}{2\pi}:\psi^\dagger_{2L}\psi_{2L}:
~~,~~
\pi_{1R}=\int
\tfrac{d\sigma}{2\pi}:\psi^\dagger_{2R}\psi_{2R}:
\end{eqnarray}
\end{subequations}

In the NS sector, the fermion vacuum is unique and the fermion
number is quantized as integers.   The above equations show that
the boson momentum is equal to $\pm$ the total fermion number of
the states there.

In the R-sector, the vacuum is degenerate and the vacuum states
$|---->$ have fermion numbers $(-1/2,-1/2,-1/2,-1/2)$ and
$\pi_{1L}=-1/2$, $\pi_{1R}=1/2$, $\pi_{2L}=1/2$ and $\pi_{2R}=-1/2$.
Operating fermion operators change these by integers, so that the
whole spectrum of fermion numbers and momenta are half-odd-integers.

Then, following ref.~\cite{Lee:2005ge}, we observe that the boson
momentum eigenstates which have no boson oscillators excited
correspond to fermion states where all fermion states are filled up
to a Fermi level. The fermion number of each type of fermions is
equal (up to signs) to the momentum of the corresponding boson
state.

\subsection{NS Sector}

In the NS-sector, the $\pi_L$ and $\pi_R$ are integers.  We denote
the states with these values of boson momentum (and vanishing
$Y$-momenta) and no boson oscillators excited by $|\pi_L,\pi_R>$.

We shall require the fermion representation of the boson momentum
state in bra-form.  The following are the states of the NS sector,
\begin{subequations}
\label{generallabel}
\begin{eqnarray}
\pi_L>0~,~\pi_R>0~~~~
<\pi_L,\pi_R|&=&<0|\prod_{r=\tfrac{1}{2}}^{\pi_L-\tfrac{1}{2}}
\left(\psi_{2r}^{\dagger}\psi_{1r}\right)
\prod_{r=\tfrac{1}{2}}^{\pi_R-\tfrac{1}{2}} \left( \tilde\psi_{2r}
\tilde \psi_{1r}^{\dagger}\right) \\
\pi_L>0~,~\pi_R<0~~~~
<\pi_L,\pi_R|&=&<0|\prod_{r=\tfrac{1}{2}}^{\pi_L-\tfrac{1}{2}}
\left(\psi_{2r}^{\dagger}\psi_{1r}\right)
\prod_{r=\tfrac{1}{2}}^{-\pi_R-\tfrac{1}{2}} \left( \tilde
\psi_{1r}\tilde\psi_{2r}^{\dagger} \right) \\
\pi_L<0~,~\pi_R>0~~~~
<\pi_L,\pi_R| &=& <0|\prod_{r=\tfrac{1}{2}}^{-\pi_L-\tfrac{1}{2}}
\left(\psi_{1r}^{\dagger}\psi_{2r}\right)
\prod_{r=\tfrac{1}{2}}^{\pi_R-\tfrac{1}{2}} \left(\tilde\psi_{2r}
\tilde \psi_{1r}^{\dagger}\right) \\
\pi_L<0~,~\pi_R<0~~~~
<\pi_L,\pi_R|&=&<0|\prod_{r=\tfrac{1}{2}}^{-\pi_L-\tfrac{1}{2}}
\left(\psi_{1r}^{\dagger}\psi_{2r}\right)
\prod_{r=\tfrac{1}{2}}^{-\pi_R-\tfrac{1}{2}} \left( \tilde
\psi_{1r}\tilde\psi_{2r}^{\dagger} \right).
\end{eqnarray}
\end{subequations}

The phases of these states are fixed by their overlaps with $|DD>$
and $|ND>$ which, recalling the known forms of these states in
boson variables, must all be either zero or one.

The boundary state in the NS sector is given in
eq.~(\ref{nsboundarystate}) which we copy here for convenience of
the reader:
\begin{equation}
|B,D>_{NS} = \tfrac{1}{\sqrt{2}} \prod_{r=\tfrac{1}{2}}^\infty
\exp\left[\psi_{-r}^\dagger U^{-1}i\sigma^1 \tilde\psi_{-r}
-\tilde\psi_{-r}^{\dagger}i\sigma^1U\psi_{-r}\right]|0>.
\end{equation}

Then, we can take the matrix elements of the momentum states and
the boundary state to obtain:

\begin{eqnarray}
\pi_L>0,\pi_R>0
\nonumber \\
<\pi_L,\pi_R|B,D>_{NS} &=&\tfrac{1}{\sqrt{2}} <0|
\prod_{r=\tfrac{1}{2}}^{\pi_L-\tfrac{1}{2}}
\left(\psi_{2r}^{\dagger}
+\tilde\psi_{a,-r}^{\dagger}[i\sigma^1U]_{a2}\right)
\left(\psi_{1r}+[U^{-1}i\sigma^1]_{1a}\tilde\psi_{a,-r}\right)  \cdot \nonumber \\
&&\prod_{r=\tfrac{1}{2}}^{\pi_R-\tfrac{1}{2}}
\left(\tilde\psi_{2r} -[i\sigma^1U]_{2a}\psi_{a,-r}\right) \left(
\tilde
\psi_{1r}^{\dagger}-\psi_{a,-r}^{\dagger}[U^{-1}i\sigma^1]_{a1}\right) |0> \nonumber \\
&=& \tfrac{1}{\sqrt{2}}
\left(-iU_{12}\right)^{2\pi_L}\delta(\pi_L-\pi_R)
\end{eqnarray}

\begin{eqnarray} \pi_L>0,\pi_R<0
\nonumber \\
<\pi_L,\pi_R|B,D>_{NS} &=& \tfrac{1}{\sqrt{2}} <0|
\prod_{r=\tfrac{1}{2}}^{\pi_L-\tfrac{1}{2}}
\left(\psi_{2r}^{\dagger}
+\tilde\psi_{a,-r}^{\dagger}[i\sigma^1U]_{a2}\right)\left(\psi_{1r}+[U^{-1}i\sigma^1]_{1a}\tilde\psi_{a,-r}\right)
\cdot \nonumber \\
&& \prod_{r=\tfrac{1}{2}}^{-\pi_R-\tfrac{1}{2}} \left( \tilde
\psi_{1r} -
[i\sigma^1U]_{1a}\psi_{a,-r}\right)\left(\tilde\psi_{2r}^\dagger
-\psi_{a,-r}^\dagger [U^{-1}i\sigma^1]_{a2}\right) |0> \nonumber \\
&=&\tfrac{1}{\sqrt{2}}
\left(U_{11}\right)^{2\pi_L}\delta(\pi_L+\pi_R)
\end{eqnarray}

\begin{eqnarray} \pi_L<0,\pi_R>0
\nonumber \\
<\pi_L,\pi_R|B,D>_{NS} &=&  \tfrac{1}{\sqrt{2}}<0|
\prod_{r=\tfrac{1}{2}}^{-\pi_L-\tfrac{1}{2}}
\left(\psi_{1r}^{\dagger} +
\tilde\psi_{a,-r}^{\dagger}[i\sigma^1U]_{a1}\right)\left(\psi_{2r}
+[U^{-1}i\sigma^1]_{2a}\tilde\psi_{a,-r}\right)\cdot \nonumber \\
&&\prod_{r=\tfrac{1}{2}}^{\pi_R-\tfrac{1}{2}}
\left(\tilde\psi_{2r} -[i\sigma^1U]_{2a}\psi_{a,-r}\right)\left(
\tilde
\psi_{1r}^{\dagger}-\psi_{a,-r}^{\dagger}[U^{-1}i\sigma^1]_{a1}\right) |0> \nonumber \\
&=&  \tfrac{1}{\sqrt{2}}\left(
U_{22}\right)^{-2\pi_L}\delta(\pi_L+\pi_R)
\end{eqnarray}

\begin{eqnarray} \pi_L<0,\pi_R<0
\nonumber \\
<\pi_L,\pi_R|B,D>_{NS} &=&  \tfrac{1}{\sqrt{2}}<0|
\prod_{r=\tfrac{1}{2}}^{-\pi_L-\tfrac{1}{2}}
\left(\psi^{\dagger}_{1r}+\tilde\psi^{\dagger}_{a,-r}[i\sigma^1U]_{a1}\right)
 \left(\psi_{2r}
+[U^{-1}i\sigma^1]_{2a}\tilde\psi_{a,-r}\right)\cdot \nonumber \\
&&\prod_{r=\tfrac{1}{2}}^{-\pi_R-\tfrac{1}{2}} \left( \tilde
\psi_{1r} -
[i\sigma^1U]_{1a}\psi_{a,-r}\right)\left(\tilde\psi_{2r}^\dagger
-\psi_{a,-r}^\dagger [U^{-1}i\sigma^1]_{a2}\right) |0> \nonumber \\
&=&  \tfrac{1}{\sqrt{2}}\left(
-iU_{21}\right)^{-2\pi_L}\delta(\pi_L-\pi_R)
\end{eqnarray}

This summarizes the overlap of momentum eigenstates with the
boundary state in the NS sector.

\subsection{R-sector}

In the R-sector, the momenta are half-odd-integral and the
momentum states of the $X$-boson are
\begin{subequations}
\label{generallabel}
\begin{eqnarray}
\pi_L>0,~\pi_R>0,
<\pi_L,\pi_R|&=&<+--+|\prod_{n=1}^{\pi_L-\tfrac{1}{2}}
\left(\psi_{2n}^{\dagger}\psi_{1n}\right)
\prod_{n=1}^{\pi_R-\tfrac{1}{2}} \left(\tilde\psi_{2n}
 \tilde
\psi_{1n}^{\dagger}\right) \\
\pi_L>0,~\pi_R<0,
<\pi_L,\pi_R|&=&i<++--|\prod_{n=1}^{\pi_L-\tfrac{1}{2}}
\left(\psi_{2n}^{\dagger}\psi_{1n}\right)
\prod_{n=1}^{-\pi_R-\tfrac{1}{2}} \left( \tilde
\psi_{1n}\tilde\psi_{2n}^{\dagger} \right) \\
\pi_L<0,~\pi_R>0,
<\pi_L,\pi_R|&=&-i<--++|\prod_{n=1}^{-\pi_L-\tfrac{1}{2}}
\left(\psi_{1n}^{\dagger}\psi_{2n}\right)
\prod_{n=1}^{\pi_R-\tfrac{1}{2}} \left( \tilde\psi_{2n}
\tilde\psi_{1n}^{\dagger}\right) \\
\pi_L<0,~\pi_R<0,
<\pi_L,\pi_R|&=&<-++-|\prod_{n=1}^{-\pi_L-\tfrac{1}{2}}
\left(\psi_{1n}^{\dagger}\psi_{2n}\right)
\prod_{n=1}^{-\pi_R-\tfrac{1}{2}} \left( \tilde
\psi_{1n}\tilde\psi_{2n}^{\dagger} \right).
\end{eqnarray}
\end{subequations}

Here, again, the phases have been adjusted so that the matrix
elements of these states with the $|DD>$ and $|ND>$ are always
either one or zero, just as the phases of the boson states
(\ref{bstates1})-(\ref{bstates2}) (with $A=0$) with boson momentum
states would be.

The R-sector boundary state is given in eq.~(\ref{rboundarystate})
which we copy here for the reader's convenience:
\begin{equation}
|BD>_R=2^{-\tfrac{1}{2}}\prod_{n=1}^\infty  \exp\left[
\psi_{-n}^\dagger U^{-1}i\sigma^1\tilde\psi_{-n} -
\tilde\psi^{\dagger}_{-n}i\sigma^1U\psi_{-n}\right]\exp\left[\psi_0^\dagger
U^{-1}i\sigma^1\tilde\psi_0\right]~|-+-+>.
\end{equation}

Then, the matrix elements are taken as
\begin{eqnarray} \pi_L>0,\pi_R>0
\nonumber \\
<\pi_L,\pi_R|B,D>_{R} &=& \frac{1}{\sqrt{2}}<+--+|
\prod_{n=1}^{\pi_L-\tfrac{1}{2}} \left(\psi_{2,n}^{\dagger}
+\tilde\psi_{a,-n}^{\dagger}[i\sigma^1U]_{a2}\right)
\left(\psi_{1,n}+[U^{-1}i\sigma^1]_{1a}\tilde\psi_{a,-n}\right)\cdot \nonumber \\
&& \quad \prod_{n=1}^{\pi_R-\tfrac{1}{2}}\left(\tilde\psi_{2,n}
-[i\sigma^1U]_{2a}\psi_{a,-n}\right) \left( \tilde
\psi_{1,n}^{\dagger}-\psi_{a,-n}^{\dagger}[U^{-1}i\sigma^1]_{a1}\right)\nonumber\\
&& \quad \exp\left[\psi_0^\dagger
U^{-1}i\sigma^1\tilde\psi_0\right]~|-+-+> \nonumber \\
&=& \frac{1}{\sqrt{2}}\left(
-iU_{12}\right)^{2\pi_L}\delta(\pi_L-\pi_R),
\end{eqnarray}
\begin{eqnarray} \pi_L>0,\pi_R<0
\nonumber \\
<\pi_L,\pi_R|B,D>_{R} &=& \frac{i}{\sqrt{2}}<++--|
\prod_{n=1}^{\pi_L-\tfrac{1}{2}} \left(\psi_{2,n}^{\dagger}
+\tilde\psi_{a,-n}^{\dagger}[i\sigma^1U]_{a2}\right)
\left(\psi_{1,n}+[U^{-1}i\sigma^1]_{1a}\tilde\psi_{a,-n}\right)\cdot \nonumber \\
&& \quad \prod_{n=1}^{-\pi_R-\tfrac{1}{2}} \left( \tilde \psi_{1,n} -
[i\sigma^1U]_{1a}\psi_{a,-n}\right)\left(\tilde\psi_{2,n}^\dagger
-\psi_{a,-n}^\dagger
[U^{-1}i\sigma^1]_{a2}\right)\nonumber\\
&& \quad \exp\left[\psi_0^\dagger
U^{-1}i\sigma^1\tilde\psi_0\right]~|-+-+> \nonumber \\
&=&\frac{1}{\sqrt{2}} \left(
U_{22}\right)^{2\pi_L}\delta(\pi_L+\pi_R),
\end{eqnarray}
\begin{eqnarray} \pi_L<0,\pi_R>0
\nonumber \\
<\pi_L,\pi_R|B,D>_{R} &=& -\frac{i}{\sqrt{2}}<--++|
\prod_{n=1}^{-\pi_L-\tfrac{1}{2}} \left(\psi_{1,n}^{\dagger} +
\tilde\psi_{a,-n}^{\dagger}[i\sigma^1U]_{a1}\right)\left(\psi_{2r}
+ [U^{-1}i\sigma^1]_{2a}\tilde\psi_{a,-n}\right)
\cdot \nonumber \\
&& \quad \prod_{n=1}^{\pi_R-\tfrac{1}{2}}\left(\tilde\psi_{2,n}
-[i\sigma^1U]_{2a}\psi_{a,-n}\right) \left( \tilde
\psi_{1,n}^{\dagger}-\psi_{a,-n}^{\dagger}[U^{-1}i\sigma^1]_{a1}\right)\nonumber\\
&& \quad \exp\left[\psi_0^\dagger
U^{-1}i\sigma^1\tilde\psi_0\right]~|-+-+> \nonumber \\
&=&\frac{1}{\sqrt{2}}\left(
U_{11}\right)^{-2\pi_L}\delta(\pi_L+\pi_R),
\end{eqnarray}
\begin{eqnarray} \pi_L<0,\pi_R<0
\nonumber \\
<\pi_L,\pi_R|B,D>_{R} &=& \frac{1}{\sqrt{2}}<-++-|
\prod_{n=1}^{-\pi_L-\tfrac{1}{2}} \left(\psi_{1,n}^{\dagger} +
\tilde\psi_{a,-n}^{\dagger}[i\sigma^1U]_{a1}\right)\left(\psi_{2r}
+ [U^{-1}i\sigma^1]_{2a}\tilde\psi_{a,-n}\right)\cdot \nonumber \\
&& \quad \prod_{n=1}^{-\pi_R-\tfrac{1}{2}} \left( \tilde \psi_{1,n}
-[i\sigma^1U]_{1a}\psi_{a,-n}\right)\left(\tilde\psi_{2,n}^\dagger
-\psi_{a,-n}^\dagger [U^{-1}i\sigma^1]_{a2}\right) \nonumber\\
&& \quad \exp\left[\psi_0^\dagger
U^{-1}i\sigma^1\tilde\psi_0\right]~|-+-+> \nonumber \\
&=& \frac{1}{\sqrt{2}}\left(
-iU_{21}\right)^{-2\pi_L}\delta(\pi_L-\pi_R).
\end{eqnarray}

We see that the overlap of momentum states with R states has a
very similar form to the overlaps with NS states.

\section{Conclusions}

 The critical theory which we have discussed in this Paper should
 be difficult to produce in a realistic system.  The reason for
 this is that, when the parameters are tuned so that precisely the
 exactly marginal boundary operator is allowed by the periodicity
 of the boson, a bulk operator with the same periodicity is
 allowed in the bulk.  Moreover, $:e^{iX}:$ is a relevant
 perturbation in the bulk, it drives a Kosterlitz Thouless
 transition which gaps the spectrum there, nullifying our
 possibility of finding the critical theory on the boundary.

In string theory, this is the Hagedorn behavior.  When a
coordinate of the bosonic string is compactified and the radius is
lowered from infinity, we expect to see a phase transition when
$R=\sqrt{2}$, before we arrive at the self-dual value $R=1$.  In
the string theory, this is fixed by going to the superstring where
super-partners of the boson provide stability.  It would be
interesting to understand this better in the context of our
present analysis.

Further, although the self-dual radius $R=1$ lies in the unstable
regime, a large number of the fractional radii $R=M/N$ do not.  On
general grounds, we expect that their critical states are less
stable than the self-dual state -- they contain fewer states, so
should have smaller entropy and therefore larger free energy --
they would avoid the Kosterlitz-Thouless transition in the bulk.

The boundary sine-Gordon theories which have fractional periods
are apparently non-local.  We have presented them as theories at
the dual radius with the addition of a discrete gauge invariance.
Such theories are indeed known to arise in models of
Josephson-junction arrays \cite{ioffe1}-\cite{Doucot:2003mh} and
whether the conformal field theories with rational radii can be
realized there is an interesting question.

The fermionic boundary states are simple and easy to work with. It
should be straightforward to use them to check some important
previous computations which used other techniques. One example is
the closed string emission amplitudes for the rolling tachyon
which were computed in \cite{Lambert:2003zr}, and should easily be
accessible in the fermion picture.

\vskip 1cm

 \noindent {\bf Acknowledgement:} The authors thank Volker Schomerus
and Ian Affleck for extensive and informative discussions. This
work is supported in part by NSERC of Canada, by the Pacific
Institute for Theoretical Physics and by KOSEF of Korea through
CQUeST (Center for Quantum Space Time). The work of TL is also
supported by Kangwon National University through the Research Fund
for 2003 Faculty Research Abroad.

\end{document}